\documentclass{aa}  

\usepackage{txfonts}
\usepackage{graphicx}
\usepackage{csquotes}
\usepackage{siunitx}
\graphicspath{{./Plots/}}

\begin{document}

\title{The CARMENES search for exoplanets around M dwarfs} 
\subtitle{Planet occurrence rates from a subsample of 71 stars}
\authorrunning{Sabotta et al.}
\titlerunning{Planet occurrence rates from a subsample of 71 CARMENES M dwarfs}

\author{
S.~Sabotta\inst{1,2}
\and 
M.~Schlecker\inst{3}
\and
P.~Chaturvedi\inst{1}
\and
E.\,W.~Guenther\inst{1}
\and
I.~Mu\~{n}oz Rodr\'iguez\inst{4}
\and
J.\,C.~Mu\~{n}oz~S\'anchez\inst{5,6}
\and
J.\,A.~Caballero\inst{7}
\and
Y.~Shan\inst{8}
\and
S.~Reffert\inst{2}
\and
I.~Ribas\inst{5,6}
\and
A.~Reiners\inst{8}
\and
A.\,P.~Hatzes\inst{1}
\and
P.\,J.~Amado\inst{4}
\and
H.~Klahr\inst{3}
\and
J.\,C.~Morales\inst{5,6}
\and
A.~Quirrenbach\inst{2}
\and
Th.~Henning\inst{3}
\and
S.~Dreizler\inst{8}
\and
E.~Pall\'e\inst{9,10}
\and
M.~Perger\inst{5,6}
\and
M.~Azzaro\inst{11}
\and
S.\,V.~Jeffers\inst{12}
\and
A.~Kaminski\inst{2}
\and
M.~K\"urster\inst{3}
\and
M.~Lafarga\inst{5,6,13}
\and
D.~Montes\inst{14}
\and
V.\,M.~Passegger\inst{15,16}
\and
M.~Zechmeister\inst{8}
}
\institute{
Th\"uringer Landessternwarte Tautenburg, Sternwarte 5, 07778 Tautenburg, Germany \\ \email{silvia.sabotta@lsw.uni-heidelberg.de}
 \and
Landessternwarte, Zentrum für Astronomie der Universit\"at Heidelberg, K\"onigstuhl 12, 69117 Heidelberg, Germany
 \and
Max-Planck-Institut f\"ur Astronomie, K\"onigstuhl 17, 69117 Heidelberg, Germany
 \and
Instituto de Astrofísica de Andaluc\'ia (IAA-CSIC), Glorieta de la Astronom\'ia s/n, 18008 Granada, Spain
 \and 
Institut de Ci\`encies de l’Espai (ICE, CSIC), Campus UAB, c/ de Can Magrans s/n, 08193 Cerdanyola del Vall\`es, Barcelona, Spain
 \and
Institut d’Estudis Espacials de Catalunya (IEEC), c/ Gran Capit\`a 2-4, 08034 Barcelona, Spain
 \and
Centro de Astrobiolog\'ia (CSIC-INTA), ESAC, Camino bajo del castillo s/n, 28692 Villanueva de la Cañada, Madrid, Spain
 \and
Institut f\"ur Astrophysik, Georg-August-Universit\"at, Friedrich-Hund-Platz 1, 37077 G\"ottingen, Germany
\and 
Instituto de Astrof\'isica de Canarias, V\'ia L\'actea s/n, 38205 La Laguna, Tenerife, Spain
\and 
Departamento de Astrof\'isica, Universidad de La Laguna, 38026 La Laguna, Tenerife, Spain
\and 
Max-Planck Institute for Solar System Research Justus-von-Liebig Weg 3, 37077 Goettingen, Germany
\and 
Centro Astron\'omico Hispano-Alem\'an (CSIC-Junta de Andaluc\'ia), Observatorio  Astron\'omico  de  Calar  Alto,  Sierra  de  los  Filabres, 04550 G\'ergal, Almer\'ia, Spain
\and 
Department of Physics, University of Warwick, Gibbet Hill Road, Coventry CV4 7AL, United Kingdom
\and 
Departamento de Física de la Tierra y Astrofísica  \& IPARCOS-UCM (Instituto de F\'isica de Part\'iculas y del Cosmos de la UCM), Facultad de Ciencias F\'isicas, Universidad Complutense de Madrid, 28040 Madrid, Spain
\and 
Hamburger Sternwarte, Gojenbergsweg 112, 21029 Hamburg, Germany
\and
Homer L. Dodge Department of Physics and Astronomy, University of Oklahoma, 440 West Brooks Street, Norman, OK 73019, United States of America
\\
}
\date{Received 31 March 2021 / Accepted 29 June 2021}

  \abstract 
{The CARMENES exoplanet survey of M dwarfs has obtained more than \num{18000} spectra of 329 nearby M dwarfs over the past five years as part of its guaranteed time observations (GTO) program.}
{We determine planet occurrence rates with the 71 stars from the GTO program for which we have more than 50 observations.}
{We use injection-and-retrieval experiments on the radial-velocity (RV) time series to measure detection probabilities. We include 27 planets in 21 planetary systems in our analysis.}
{We find $0.06^{+0.04}_{-0.03}$ giant planets ($\SI{100}{M_\oplus} < M_\text{pl} \sin i < \SI{1000}{M_\oplus}$) per star in periods of up to \SI{1000}{d}, but due to a selection bias this number could be up to a factor of five lower in the whole 329-star sample. The upper limit for hot Jupiters (orbital period of less than 10\,d) is 0.03 planets per star, while the occurrence rate of planets with intermediate masses ($\SI{10}{M_\oplus} < M_\text{pl} \sin i < \SI{100}{M_\oplus}$) is $0.18^{+0.07}_{-0.05}$ planets per star. Less massive planets with $ \SI{1}{M_\oplus} < M_\text{pl} \sin i < \SI{10}{M_\oplus}$ are very abundant, with an estimated rate of $1.32^{+0.33}_{-0.31}$ planets per star for periods of up to 100\,d. When considering only late M dwarfs with masses $M_\star < \SI{0.34}{M_\odot}$, planets more massive than \SI{10}{M_\oplus} become rare. Instead, low-mass planets with periods shorter than 10\,d are significantly overabundant.}
{For orbital periods shorter than 100\,d, our results confirm the known stellar mass dependences from the \emph{Kepler} survey: M dwarfs host fewer giant planets and at least two times more planets with $M_\text{pl}~\sin i < \SI{10}{M_\oplus}$ than G-type stars. In contrast to previous results, planets around our sample of very low-mass stars have a higher occurrence rate in short-period orbits of less than 10\,d. Our results demonstrate the need to take into account host star masses in planet formation models.
}
 \maketitle

\section{Introduction }
\label{sec:introduction}

An open question in exoplanet research is if planet formation theories produce the same exoplanet population as the observed one. In order to answer this question, it is crucial to study the population of planets as a function of stellar mass. We expect the planet occurrence rate to change with the mass of the host star because more massive stars host more massive protoplanetary disks \citep{Mordasini2012a,Andrews2013,Pascucci2016,Ansdell2016b,Tychoniec2018}.

Exoplanet surveys of low-mass M dwarfs are very important because they investigate the most abundant type of star \citep{Henry2006, Henry2018, Reyle2021}. They have the potential to detect small, low-mass planets (of a few Earth radii and masses) because the radial-velocity (RV) signal of low-mass planets and the transit depth of small planets orbiting this type of star are larger than for solar-like stars. 

The first piece of evidence that the planet population around M dwarfs is different from that of hotter stars is provided by \cite{Endl2006}, who shows that hot Jupiters ($P < 10$\,d and $M\sin i$ > \SI{100}{M_\oplus}) are rare around M dwarfs compared to solar-like stars. This realization withstands the test of time. Despite favorable sensitivity to hot Jupiters, M dwarf planet surveys result in very few such discoveries \citep{Johnson2012, Hartman2015, Bayliss2018, Bakos2020}. On the other hand, the sample size of the M dwarf surveys that are conducted thus far is not large enough to individually rule out a hot Jupiter occurrence rate that is consistent with that of hot Jupiters around G dwarfs \citep[e.g.,][]{Obermeier2016}.

Although the first planets orbiting M dwarfs are discovered by RV surveys \citep{Marcy2001,Butler2004,Bonfils2005}, a big leap forward is the \emph{Kepler} satellite \citep{Borucki2010}, which discovers thousands of transiting planets and candidates and shows that the frequency of small planets around low-mass stars, at least in short periods on the order of 100\,d, is higher than that around solar-like stars \citep{Howard2012}. Subsequent studies of the \emph{Kepler} sample look at either planet occurrence rates specifically for the M dwarf sample  \citep[such as][]{Dressing2013,Dressing2015,Hsu2020} or the stellar mass dependence throughout the full \emph{Kepler} sample \citep[such as][]{Mulders2015, Yang2020}. The survey has a very strong focus on G-type stars. Therefore, the size of the M dwarf sample is much lower than that of the G dwarf sample, and the lowest-mass M dwarfs are underrepresented. As an example, only 1.5\,\% (58 stars) of the M dwarfs investigated by \cite{Dressing2013} have a mass below \SI{0.15}{M_\odot}.
Any study of a mass dependence of planet occurrence rates from the \emph{Kepler} M dwarf sample thus comes with large error bars \citep{HardegreeUllman2019}.

An advantage of RV surveys is that they can detect all planets, transiting or not, that induce a sufficiently large velocity variation in their host star. The RV amplitude of a super-Earth planet ($M \sim$ 3--5\,M$_\oplus$) in the habitable zone of an M dwarf is on the order of 3 to 5\,m\,s$^{-1}$, which opens up the thrilling possibility of detecting terrestrial planets orbiting the nearest stars. Nevertheless, the analysis of the M dwarf sample of HARPS (High Accuracy Radial velocity Planet Searcher) by \citet{Bonfils2013} is still the only large statistical analysis of an RV survey of planets around this type of star. 

M dwarfs emit most of their light at near-infrared wavelengths. Therefore, red-sensitive optical and infrared spectrographs are better suited for exoplanet surveys than spectrographs working at bluer wavelengths. 
At these spectral types, the sweet spot with the highest RV precision is the red part of the optical regime and the near infrared, from \SIrange{700}{900}{nm} \citep{Reiners2018a}. This consideration leads to the construction of CARMENES \citep[Calar Alto high-Resolution search for M dwarfs with Exo-earths with Near-infrared and optical Echelle Spectrographs;][]{Quirrenbach2014}. CARMENES is a stabilized, fiber-fed, two-channel echelle spectrograph with resolution $R\sim$ \num{80000}--\num{95000} that covers, in one shot, the wavelength region between \SI{520}{nm} and \SI{1710}{nm}. The instrument is operational for over five years and, through its systematic monitoring program, discovers a large number of planets around nearby M dwarfs (see below). As a result, the CARMENES survey is well suited to determine the frequency of planets around low-mass stars.
The CARMENES survey is currently unique because of the large amount of invested observing time and due to it being the first survey to use such a large wavelength range.
The CARMENES consortium is awarded 750 useful nights as guaranteed time observations (GTO), which are conducted between January 2016 and December 2020.
After its recent completion, the consortium starts a new legacy RV exoplanet survey with a comparable amount of awarded time, which is essentially a continuation of the GTO program. 
In contrast to previous surveys that use spectrographs in the optical regime like ELODIE, CORALIE, HARPS and the High Resolution Echelle Spectrometer HIRES, the sensitivity of CARMENES to longer wavelengths allows it to focus on M3.0\,V to M5.0\,V stars, namely, stars that typically have half the mass of the targets studied previously. 

Statistics of low-mass planets (with $M_\text{pl} <$ \SI{10}{M_\oplus}) are not only important for determining their actual frequency, but also for testing theories of planet formation. Most of the existing theories aim to reproduce the planet population in the Solar System and around other solar-like stars and, thus, assume a single solar analog host star \citep[e.g.,][]{Ida2010, Mordasini2012, Ndugu2018, Emsenhuber2020,Schlecker2020,Schlecker2021}. Crucial new insights will emerge from the adaptation of the same underlying theoretical frameworks to planets around low-mass stars and the comparison to observational data.

In this paper we present planet occurrence rates determined from the first 71 stars for which we have more than 50 observations. 
After this introduction, we continue in Sect.~\ref{sec:data_stars} with a short description of CARMENES data in general and of the stellar sample considered in this study. 
In Sect.~\ref{sec:planets} we describe how we obtained the planet subsample that we use for further analysis. In Sect.~\ref{sec:completeness} we describe our injection-and-retrieval experiment to obtain detection limits. 
We present the method and results of our occurrence rate analysis in Sect.~\ref{sec:occurrence}. 
The results are discussed in Sect.~\ref{sec:discussion}, followed by the conclusions in Sect.~\ref{sec:conclusion}.

\section{Data}
\label{sec:data}

Our occurrence rate study is based on 6512 spectra of 71 stars obtained with CARMENES from January 2016 until March 2020. 
The instrument is located at the 3.5\,m telescope of the Calar Alto Observatory in Almer\'ia, Spain. 
The spectra went through the standard GTO data flow and were reduced with the {\tt caracal} pipeline \citep{Caballero2016b}. 
We obtained RV information with the SpEctrum Radial Velocity AnaLyser ({\tt serval}) with a precision on the 1\,m\,s$^{-1}$ level \citep{Zechmeister2018}. This precision was reached after we corrected for the nightly zero point, which was derived from a subsample of stars with small RV variability \citep{Trifonov2018}. We base our analysis on the visual channel observations, which cover the wavelength range \SIrange{520}{960}{nm}, as the RV precision of the visual channel is more suitable for planet detections \citep[but there are detections that combine data from both channels, such as ][]{Bauer2020}.

\subsection{Stellar sample}\label{sec:data_stars}

\begin{figure}
 \centering
\includegraphics[width=\linewidth]{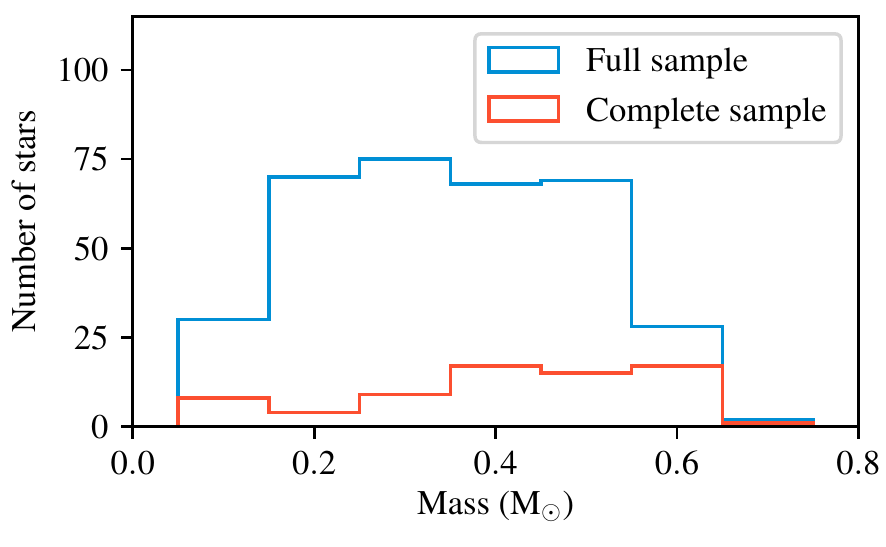}
\includegraphics[width=\linewidth]{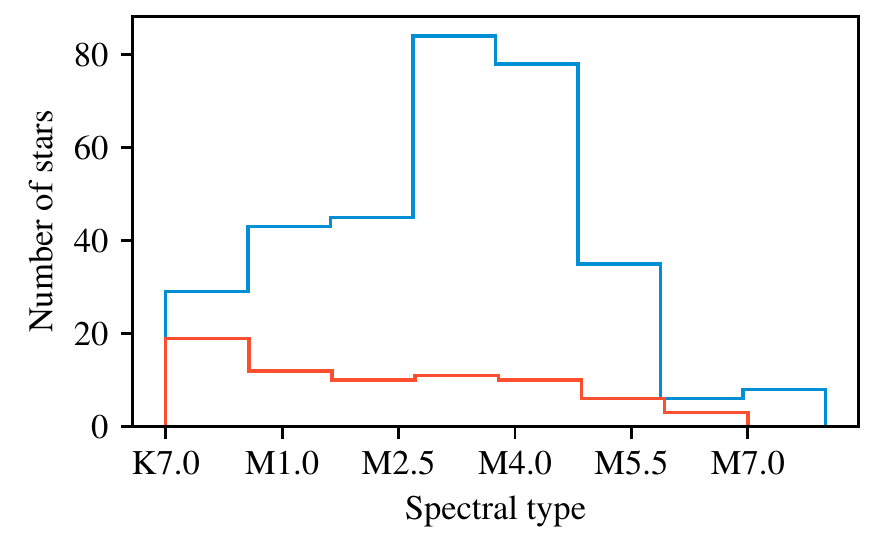}
\caption{Histograms of stellar mass ({\em top panel}) and spectral type ({\em bottom panel}) of the full GTO CARMENES sample with 329 stars (blue) and the complete subsample with 71 stars (red).}
 \label{fig:diagnostics}
\end{figure}

The 329 GTO stars of the CARMENES survey are the brightest M dwarfs for their spectral subtype in the input catalog  CARMEN(ES) Cool dwarf Information and daTa Archive \citep[Carmencita;][]{Caballero2016}. From this sample, we excluded stars that turned out to be a spectroscopic binaries. This is usually apparent after a few observations \citep{Baroch2018}. Moreover, we excluded all those stars that were added to the GTO program later, such as the Transiting Exoplanet Survey Satellite {\em (TESS)} objects of interest \citep[e.g.,][]{Bluhm2020, Dreizler2020, Kemmer2020}, as including them would bias our occurrence rate study. We also excluded very active targets with an RV scatter of more than \SI{10}{m\,s^{-1}} and $v\sin i >$\,\SI{2}{km\,s^{-1}}, namely \enquote{RV-loud} targets \citep[see][]{TalOr2018}. From the remaining sample, we selected the first 71 stars with at least 50 observations, as we do not intend to observe them further.

Figure~\ref{fig:diagnostics} shows histograms of the stellar mass ($M_\star$) and spectral type of the full sample (329 stars) and the complete subsample (71 stars). 
The masses of 315 stars in the top panel are taken from \cite{Schweitzer2019}.
The masses of six stars that are not in the complete sample are calculated with the mass-luminosity-metallicity relation from \cite{Mann2019}. In addition, eight stellar masses are dynamical masses from binary stars, which are excluded from our 71 star sample \citep{Baroch2018}. 
The median mass of the whole CARMENES sample is 0.348\,M$_\odot$ and that of our subsample is 0.426\,M$_\odot$. Thus, stars more massive than 0.348\,M$_\odot$ are over-represented in our subsample. This is mainly because we excluded the very active RV-loud stars that also have a high RV scatter. Those are more often of lower mass \citep{TalOr2018}.

A histogram of spectral types is shown in the bottom panel of Fig.~\ref{fig:diagnostics}. 
The spectral type distribution in our complete sample is relatively homogeneous from M0.0\,V to M6.0\,V (with one K7.0\,V target), whereas the full CARMENES GTO sample concentrates more on the spectral types from M3.0\,V to M5.0\,V.  

\subsection{CARMENES planets}\label{sec:planets}

Several of the published CARMENES planets are discovered in combination with data from other instruments such as HARPS at the European Southern Observatory (ESO) 3.6\,m Telescope or HIRES at the Keck I Telescope \citep[e.g., \object{Barnard's Star\,b};][]{Ribas2018}. 
Since the detection limits are calculated only for the CARMENES survey, our occurrence rate analysis should be based on a planet sample detectable purely from CARMENES data. 

\begin{table}
\caption{Planets used for occurrence rate calculation.}       
\label{table:planets1mchapter}    
\centering     \onehalfspacing                           
\begin{tabular}{l l c c c l}      
\hline\hline                     
\noalign{\smallskip}
Karmn & ~ & $P$ & $M_\text{pl} \sin i$ & RV FAP & Ref.  \\  
~ & ~ & (d) & (\si{M_\oplus}) & (\%) & ~ \\
\noalign{\smallskip}
\hline   
\noalign{\smallskip}
J01125--169         & c &   3.060   &  $1.14^{+0.11}_{-0.10}$  &   0.0047 & Sto20a \\ 
J01125--169         & d &   4.656   &  $1.09^{+0.12}_{-0.12}$  &   0.0349 & Sto20a \\ 
J02530+168          & b &   4.910   &  $1.05^{+0.13}_{-0.12}$  &   $< 10^{-6}$ & Zec19 \\ 
J02530+168          & c &   11.41   &  $1.11^{+0.16}_{-0.15}$  &   $< 10^{-6}$ & Zec19 \\ 
J03133+047          & b &   2.291   &  $3.95^{+0.42}_{-0.43}$  &   $< 10^{-6}$  & Bau20 \\
J06548+332          & b  &   14.24   & $4.00^{+0.40}_{-0.40}$   &   $< 10^{-6}$ &Sto20b \\
J08413+594          & b &   203.6  &  $147^{+7.0}_{-7.0}$      &   $< 10^{-6}$ & Mor19\\
J09144+526          & b &   24.45    &  $10.3^{+1.5}_{-1.4}$   &   0.0012 & Gon20 \\
J11033+359          & b &   12.95   &  $2.69^{+0.25}_{-0.25}$    &  $< 10^{-6}$ & Sto20b \\
J11417+427          & b &   41.38   &  $96.7^{+1.4}_{-1.0}$    &  $< 10^{-6}$ & Tri18 \\
J11417+427          & c &   532.5  &  $68.1^{+4.9}_{-2.2}$  &  $< 10^{-6}$ &  Tri18\\
J11421+267          & b &   2.644    &  $21.4^{+0.20}_{-0.21}$   & $< 10^{-6}$  & Tri18 \\
J12123+544S         & b &   13.67   &  $6.89^{+0.92}_{-0.95}$    &  $< 10^{-6}$ & Sto20b \\
J12479+097$^a$      & b &   1.467    &  $2.82^{+0.11}_{-0.12}$    &   0.0012  & Tri21\\
J13229+244          & b &   3.023    &  $8.0^{+0.5}_{-0.5}$    & $< 10^{-6}$  & Luq18 \\
J16167+672S         & b &   86.54    &  $24.7^{+1.8}_{-2.4}$   & $< 10^{-6}$  & Rei18 \\
J16303--126$^b$     & b &   1.26    &  $1.92^{+0.37}_{-0.37}$    &   0.453 & Wri16\\
J16303--126         & c &   17.87   &  $4.15^{+0.37}_{-0.37}$    &   0.0011 & Wri16\\
J17378+185          & b &   15.53   &  $6.24^{+0.58}_{-0.59}$  &  0.0007 & Lal19\\
J19169+051N         & b &   105.9  &  $12.2^{+1.0}_{-1.4}$   &  $< 10^{-6}$ & Kam18\\
J21164+025          & b &   14.44   &  $13.3^{+1.0}_{-1.1}$   &  $< 10^{-6}$ & Lal19\\
J21466+668          & b &   2.305   &  $2.50^{+0.29}_{-0.30}$ &  $< 10^{-6}$ & Ama21\\
J21466+668          & c &   8.052    &  $3.75^{+0.48}_{-0.47}$ &  $< 10^{-6}$ & Ama21\\
J22137--176         & b &   3.651    &  $7.4^{+0.5}_{-0.5}$    &  $< 10^{-6}$ & Luq18 \\
J22252+594          & b &   13.35   &  $16.6^{+0.94}_{-0.95}$   &  $< 10^{-6}$ & Nag19\\
J22532--142         & b &   61.08   &  $242^{+0.7}_{-0.7}$ &  $< 10^{-6}$ & Tri18\\
J22532--142         & c &   30.13   &  $761^{+1.0}_{-1.0}$    &  $< 10^{-6}$ & Tri18\\ 
\noalign{\smallskip}
\hline                                         
\end{tabular}\singlespacing
\tablebib{
    Ama21: \citealt{Amado2021};
    Bau20: \citealt{Bauer2020};
    Gon20: \citealt{GonzalezAlvarez2020};
    Kam18: \citealt{Kaminski2018};
    Lal19: \citealt{Lalitha2019};
    Luq18: \citealt{Luque2018};
    Mor19: \citealt{Morales2019};
    Nag19: \citealt{Nagel2019};
    Rei18: \citealt{Reiners2018};
    Sto20a: \citealt{Stock2020};    
    Sto20b: \citealt{Stock2020a};
    Tri18: \citealt{Trifonov2018};
    Tri21: \citealt{Trifonov2021};
    Wri16: \citealt{Wright2016};
    Zec19: \citealt{Zechmeister2019}.\\
    \tablefoottext{a}{We tabulate the actual mass of the transiting planet GJ\,486 (J12479+097).}
    \tablefoottext{b}{The orbital period of GJ\,628\,b (J16303--126) reported by Wri16 was 4.89\,d.}
}
\end{table}

The experience from the \emph{Kepler} survey shows that the criteria that are applied for the detection limit method need to be the same to those applied to detect planets \citep{Gaudi2021}. Furthermore, for the computation of an unbiased occurrence rate, no data from other surveys should be used \citep[cf.][]{Gaudi2021}. If we include planets from other surveys we do not have the information on the detection limits and survey completeness and we cannot correct for missing planets. We would over-correct the occurrence rates because the planets are below the detection limit of our survey alone (see also Sect.~\ref{sec:discussion}). Therefore, we need to identify those planets that we can include in our analysis independently. 

We computed generalized Lomb-Scargle (GLS) periodograms \citep{Zechmeister2009} of all 71 time series. 
All peaks with a false alarm probability (FAP) of less than 1\,\% were modeled with Keplerian orbits. 
The models were calculated with the python package {\tt PyAstronomy} \citep{Czesla2019}. 
We subtracted the orbits and looked for periodogram peaks again. 
If there was a second peak, we modeled both planet candidates with a double Keplerian. 
We repeated this procedure for up to three signals. 
Usually, one would repeat the pre-whitening process until no signals with FAP $<$ 1\,\% remain in the data. 
In our data set, though, there are several signals that cannot be removed with a Keplerian model. 
In addition, a uniform analysis of the signals becomes more challenging when more signals per star are included.

M dwarfs may be active and, therefore, excluding activity peaks is crucial to avoid identifying spurious planet candidates. 
The activity-induced RV-jitter for a certain activity level is known to be larger in M- than in G-type stars \citep[e.g.,][]{Barnes2011, Jeffers2014, Suarez2017}. 
We used four activity indicators to flag the periodogram peaks in this analysis: the H$\alpha$ index and the Ca~{\sc ii} infrared triplet (CaIRT), which are sensitive to chromospheric activity, the chromatic index (CRX), which traces the dependence of the RV amplitude on the wavelength, and the differential line width (dLW), which traces changes in the line widths and is an alternative, differential indicator to full width at half maximum. 
The definition of the four indices is given in  \cite{Zechmeister2018} and \cite{Schoefer2019}.
All indicators were computed by {\tt serval}.

We retrieved 118 periodic signals with FAP $<$ 1\,\%. 
We flagged them as either \enquote{Planet,} if they corresponded to a published planet, or as \enquote{Unsolved,} to be checked further if all of the following criteria apply (see also Table~\ref{table:planets}):
(i) The period of the signal is shorter than half of our observational time baseline. Otherwise, we could not confirm the periodic nature of this signal. Longer-period signals were flagged as \enquote{$P >$ time baseline/2.}
(ii) The signal is not present in any of the four CARMENES activity indicators. If we saw it in any of the activity indicators, we flagged it as \enquote{Activity.} (iii) The signal (or its first harmonic) is not near the rotational period tabulated by Carmencita; otherwise, it was flagged as \enquote{Rotation.} If a signal that met the period criterion had a very small FAP, $<10^{-8}$, we skipped the activity analysis and flagged the signal as Unsolved or Planet in any case because such a signal needed to be analyzed manually. 

The automated signal detection process returned 27 planets (with published data) and another 18 signals flagged as Unsolved. The Unsolved signals are most probably caused by stellar activity. We reach this conclusion because their amplitude and phase are not stable over time or because the signal is a second harmonic of the rotation period.

The 27 planets that were identified in this way are only those planets for which the CARMENES observations were already thoroughly investigated and published. 
All these planets and the corresponding CARMENES publications are listed in Table~\ref{table:planets1mchapter}. The only exception is a planetary system discovered by \cite{Wright2016} with HARPS around \object{GJ\,628} (J16303--126). 
We identify the signals of the two inner planets in our periodograms. However, we find the inner planet \object{GJ\,628\,b} at 1.26\,d -- an alias of the published period at 4.89\,d. We also obtain a higher RV amplitude of $ 2.42^{+0.39}_{-0.32}$\,\si{m\,s^{-1}} as compared to $1.67^{+0.20}_{-0.19}$\,\si{m\,s^{-1}} in a newer publication on this planet \citep{AstudilloDefru2017}. This amplitude discrepancy and ambiguity in period need more thorough investigation. We include this planet with the parameters derived from our data (see Table~\ref{table:planets}). The new period and $M_\text{pl} \sin i$  are close enough to the ones that are published such that this will not affect our occurrence rate conclusions.

\begin{figure*}
    \centering
    \includegraphics[width=\linewidth]{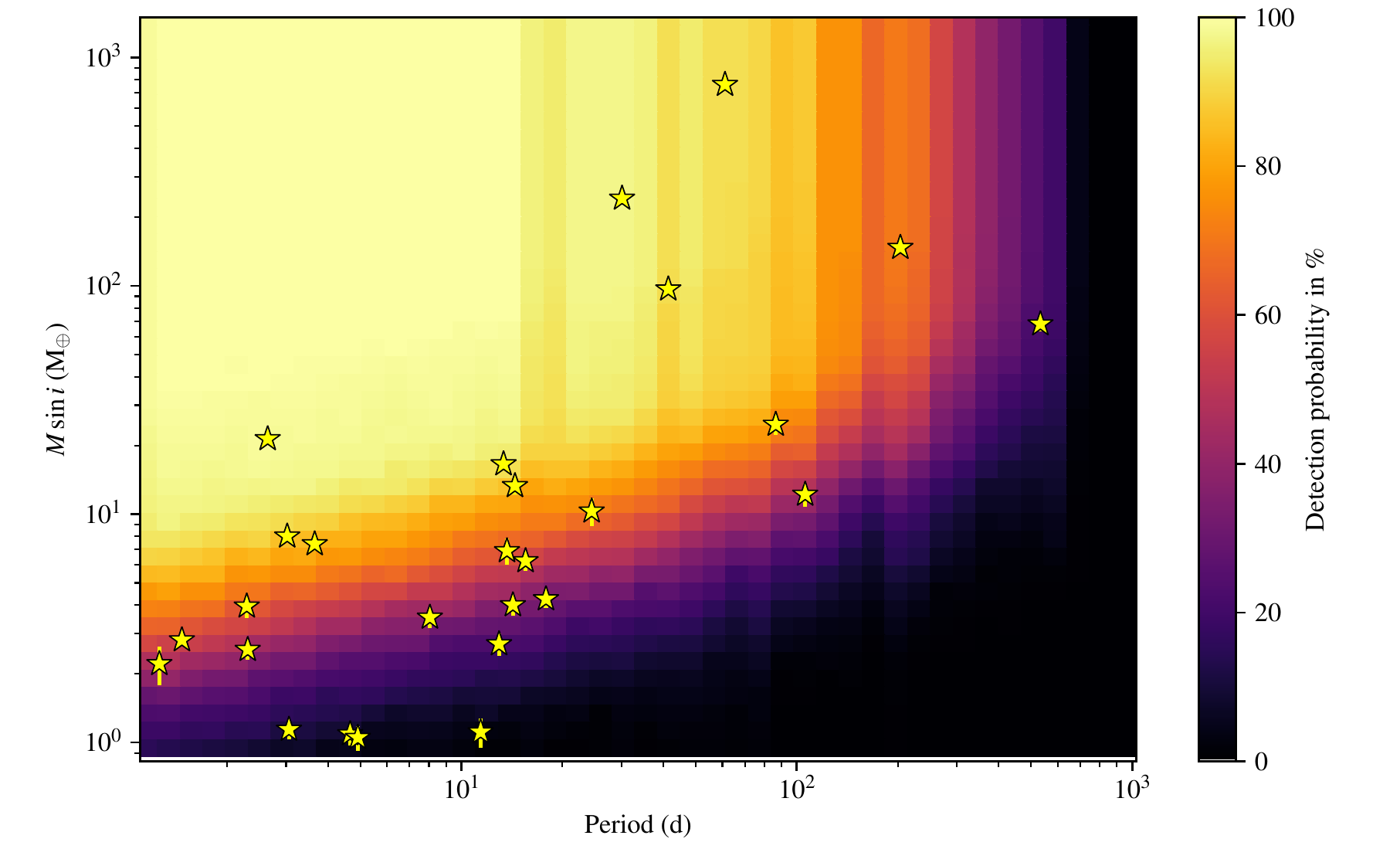}
    \caption{CARMENES GTO survey detection completeness for the subsample of 71 stars.
    The color map indicates the average detection probability of the corresponding period-mass combination.
    Yellow stars indicate planets discovered by CARMENES (error bars are sometimes smaller than the marker size).}
    \label{fig:carmenes_completeness}
\end{figure*}

\begin{figure*}
    \centering
    \begin{minipage}{0.45\textwidth}
    \includegraphics[width=\linewidth]{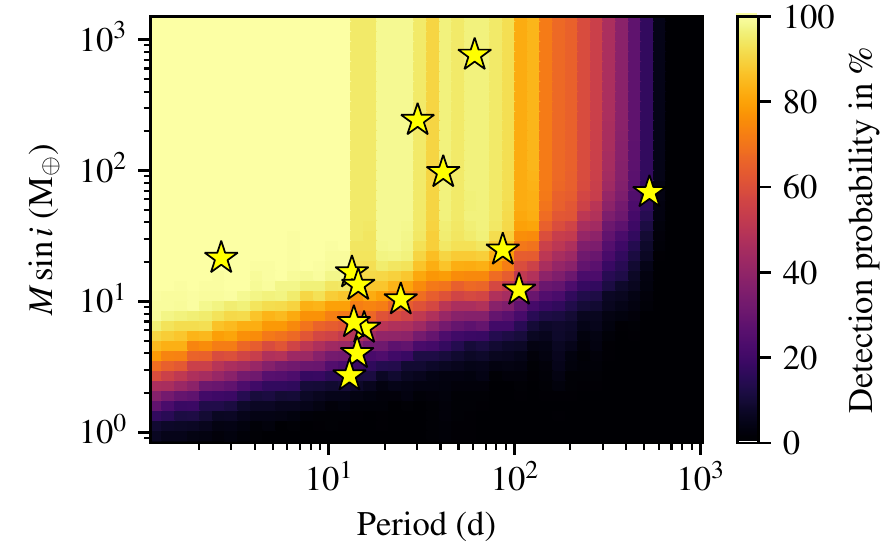}
    \end{minipage}
    \begin{minipage}{0.45\textwidth}
    \includegraphics[width=\linewidth]{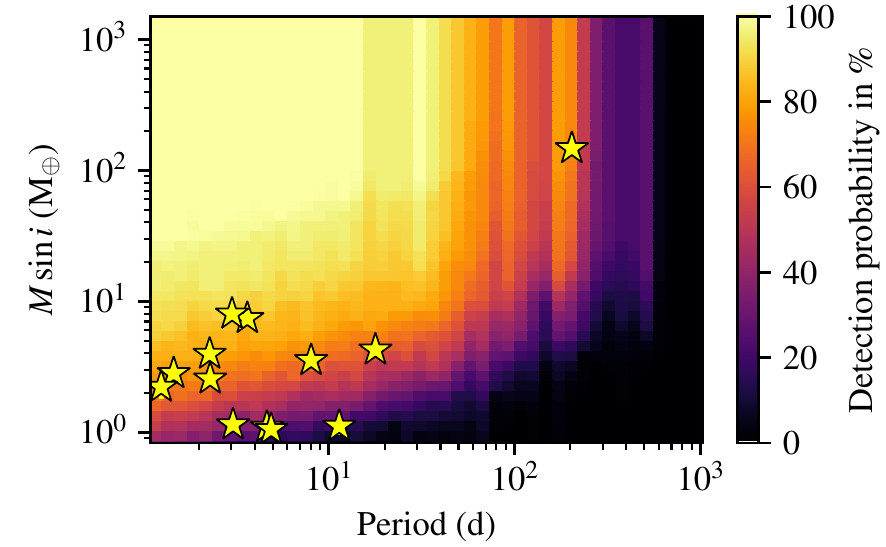}
    \end{minipage}
    \caption{Same as Fig.~\ref{fig:carmenes_completeness} but for the subsamples of 48 stars with $M_\star > \SI{0.34}{M_\odot}$ ({\em left}) and of 23 stars with $M_\star < \SI{0.34}{M_\odot}$ ({\em right}).}
    \label{fig:carmenes_completeness_high_low_mass}
\end{figure*}

\section{Analysis and results}
\subsection{Planet detection completeness}\label{sec:completeness}

The completeness of the  planet detections in our sample was calculated with an injection-and-retrieval experiment similar to other occurrence rate studies \citep[e.g.,][]{Cumming1999, Zechmeister2009, Meunier2012, Bonfils2013}. 
In particular, we injected single planets with circular orbits into our RV data and tested if we could retrieve them with a GLS periodogram (the effect of eccentric orbits is discussed in Sect.~\ref{sec:eccen_multi}).
For this purpose, we created a log-uniform grid in planet minimum mass, $M_\text{pl} \sin i$, and period, $P$, of 60 grid points each in the ranges 1--\num{10000}\,\si{M_\oplus} and 1--\num{10000}\,d, respectively, and used those mass-period combinations as parameters for our injected test planets. For circular orbits the amplitude of the simulated RV curves was computed by the approximation:
\begin{equation}\label{equ:amplitude}
K = 28.435~\si{m\,s^{-1}} \left(\frac{P}{\text{1\,yr}}\right)^{-1/3} \left( \frac{M_{\text{pl}}\sin i}{\text{M}_{\text{Jup}}}\right) \left(\frac{M_\star}{M_\odot}\right)^{-2/3}.
\end{equation}

There are two possible approaches on how to include measurement errors: we could calculate the standard deviation of the pre-whitened data set and add random noise with the same standard deviation \citep[cf.][]{Cumming1999} or inject the planet signal directly into the pre-whitened observed RV data \citep[see, e.g.,][]{Bonfils2013}. 
We used the second method since it avoids any assumption on the distribution of measurement uncertainties. 
Another advantage is that this approach preserves the stellar activity signal.
The RVs of the simulated time series are then:

\begin{equation}
RV(t) = K \sin\left(\frac{2\pi t}{P} + \phi\right) + RV_{\text{observed}}(t),
\end{equation}

\noindent where $P$ is the period of the planetary orbit, $\phi$ is a random phase angle, $RV_{\text{observed}}(t)$ is the RV at the time stamp after pre-whitening, and $t$ is the time stamp of the observation.
The RV semi-amplitude $K$ is derived from the projected mass $M_\text{pl} \sin i$ assigned to the planet (Eq.~\ref{equ:amplitude}).

We obtained a detection map as a function of $M_\text{pl} \sin i$ and $P$ for every star in our sample as follows. We simulated the RV signals of 50 test planets at each grid point ($M_\text{pl} \sin{i}_\text{i}, P_\text{i}$) with randomized phase angles,according to the prescriptions outlined above. Then, we injected each signal into the star's pre-whitened RV data and attempted to recover them. The criteria for a successful recovery were that the highest peak of the GLS periodogram has a FAP $<$ 1\,\% and is at the same period as the injected period (the tolerance is the peak width, which is the inverse of the time span in frequency space). The detection probability at every grid point is then the ratio of retrieved to injected planets: $ p_\text{pl, det,i}(M_\text{pl} \sin{i}_\text{i}, P_\text{i}) = N_\text{retrieved,i}/N_\text{injected,i}$.
The detection map of the whole survey, shown in Fig.~\ref{fig:carmenes_completeness} (and Fig.~\ref{fig:carmenes_completeness_high_low_mass} for two stellar mass bins), is then an average detection probability of every $M_\text{pl} \sin i$ and period combination.

\begin{figure}
    \centering
    \includegraphics[width=\linewidth]{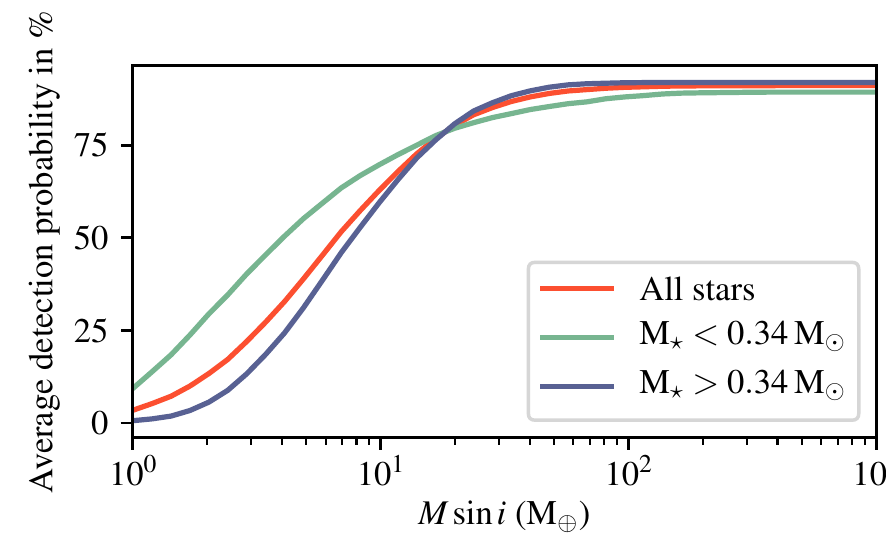}
    \caption{Average planet detection probabilities of the complete sample (subsample of 71 CARMENES GTO stars) as a function of projected planet mass averaged over the period range 1--240\,d.
    The three curves depict the probabilities for early-M (blue), late-M (green), and all stars (red).
    The maximum probability is 0.91 on average.}
    \label{fig:sensitivity_mass}
\end{figure}

Figure~\ref{fig:sensitivity_mass} presents the survey sensitivity in a different way. 
It shows that we cannot detect Earth-mass planets around stars more massive than \SI{0.34}{M_\odot}. The detection probability increases steeply for planets of \SI{2}{M_\oplus} or more. 
Around later (i.e., less massive) M dwarfs, our CARMENES RV survey is able to detect some Earth-mass planets. 
We chose $M \sin{i}$ = \SI{0.34}{M_\odot} as the dividing line between early- and late-type M dwarfs.
This mass boundary corresponds approximately to spectral type M3.5\,V and $T_\text{eff} \approx \SI{3300}{K}$ and the threshold between fully and not fully convective stars \citep{Cifuentes2020}.

\subsection{Occurrence rates}\label{sec:occurrence}

\begin{figure*}
    \centering
    \includegraphics[width=\linewidth]{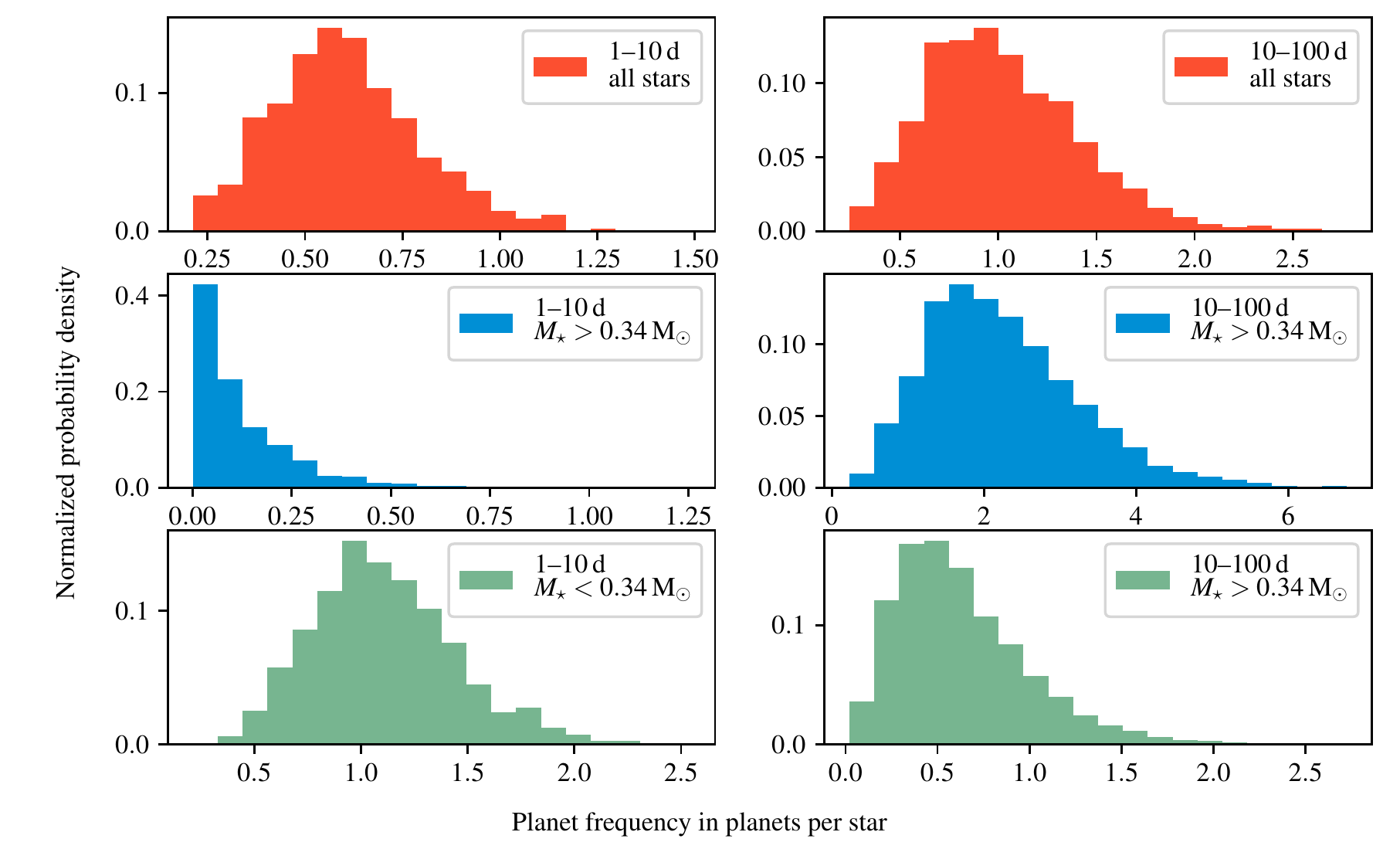}
    \caption{Output of the Monte Carlo simulation for planets with $\SI{1}{M_\oplus} < M_\text{pl} < \SI{10}{M_\oplus}$. The probability density is normalized such that the sum of all discrete points is unity.}
    \label{fig:simulation}
\end{figure*}

To finally obtain the occurrence rates, we ran a Monte Carlo simulation. For this purpose, we created a grid of test planet frequencies $\overline{n}_\text{pl}$ in number of planets per star. For each of those test frequencies, we wanted to obtain the probability that this frequency is consistent with the number of planets $N_\text{pl,det}$ that were detected in this period-mass bin.

Therefore, each simulation run followed the following four steps:
(i) We drew a number of test planets, $N_\text{pl,in}$, that corresponds to the test planet frequency from a Poisson distribution with $\lambda = \overline{n}_\text{pl}\,N_\star$ (where $N_\star$ is the number of stars in the sample). (ii) We assigned every test planet a minimum mass, $M_\text{pl} \sin i$, and orbital period, $P$, from the mass-period grid of our detection map. (iii) We accepted the test planet as a planet detection with the detection probability at the given $M_\text{pl} \sin i$ and $P$. (iv) We counted the number of test planet retrievals, $N_\text{pl,out}$.

We ran this simulation 200 times for every test frequency and counted the number of times that $N_\text{pl,out}$ was equal to $N_\text{pl,det}$. The resulting probability density was normalized such that the sum of all points was 1. We show a binned version of the simulation output in Fig.~\ref{fig:simulation}. We utilized the cumulative probability at the 16\,\%, 50\,\%, and 84\,\% levels as the lower limit, median and upper limits, respectively. Where we derived upper limits  (for bins where $N_\text{pl,det} = 0$), we took the 84\,\% level as an upper limit. This method is a variant of the method used by \cite{Bonfils2013} with the HARPS M dwarf sample.

\cite{Bonfils2013} repeatedly drew $N_\star$ random probabilities in every bin and took the 16\,\% and 84\,\% levels of the resulting distribution as error bars. If the probabilities are randomly drawn from the $\log P$-$\log M_\text{pl} \sin i$ grid, it is implicitly assumed that the planet $M_\text{pl} \sin i$ and $P$ are log-uniformly distributed. However, we know from \emph{Kepler} that this is probably true neither for the period \citep{Mulders2015} nor for the radius \cite[e.g.,][]{Foreman-Mackey2014} distributions of small planets. 
Nevertheless, we adapted the assumption of a log-uniform distribution in period and discuss why this assumption does not bias our results significantly in Sect.~\ref{sec:discussion}. For the mass, on the other hand, we used two power-law distributions, whose parameters we infer below.  

\begin{figure}
 \centering
\includegraphics[width=\linewidth]{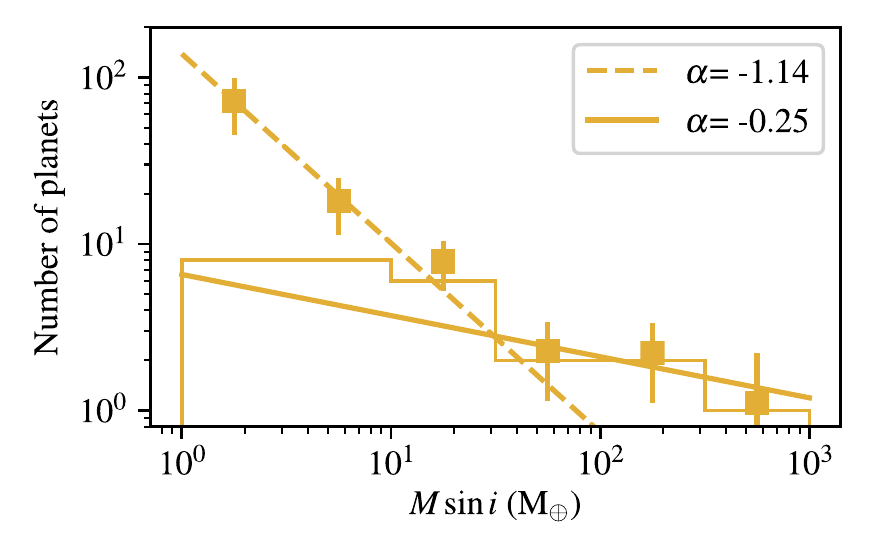}
\caption[Number of planet detections as a function of planet mass]{Number of CARMENES planet detections as a function of planet mass. 
The histogram shows the number of CARMENES planet detections and the squares the number of planets corrected for survey sensitivity. The dashed line is a power law fit to the corrected number of planets in the bins with $M_\text{pl}<\SI{32}{M_\oplus}$, and the solid line is a power law fit to the bins with $M_\text{pl}>\SI{32}{M_\oplus}$.}
\label{fig:mass_histogramm}
\end{figure}

In Fig.~\ref{fig:mass_histogramm} we show a histogram of the CARMENES planet detections. 
Every minimum mass bin of this histogram was corrected with a correction factor $C$, which is the inverse of the average detection efficiency per bin (see Fig.~\ref{fig:sensitivity_mass}). As a result, we obtained corrected numbers of planets $N_\text{pl, corr} = C ~ N_\text{pl,det}$ for six bins of $M_\text{pl} \sin i$. We fit those six data points with two power laws of the type:

\begin{equation}\label{eq:power}
    N_\text{pl} = a (M_\text{pl} \sin i)^\alpha,
\end{equation}

\noindent with the breaking point at \SI{32}{M_\oplus} (\SI{0.1}{M_\text{Jup}}). 
We measured $\alpha = -0.26\pm0.17$ for planets with $M_\text{pl} \sin i >  \SI{32}{M_\oplus}$ and $\alpha = -1.14\pm0.16$ for planets with $M_\text{pl} \sin i <  \SI{32}{M_\oplus}$.

\begin{figure}
 \centering
 \includegraphics[width=\linewidth]{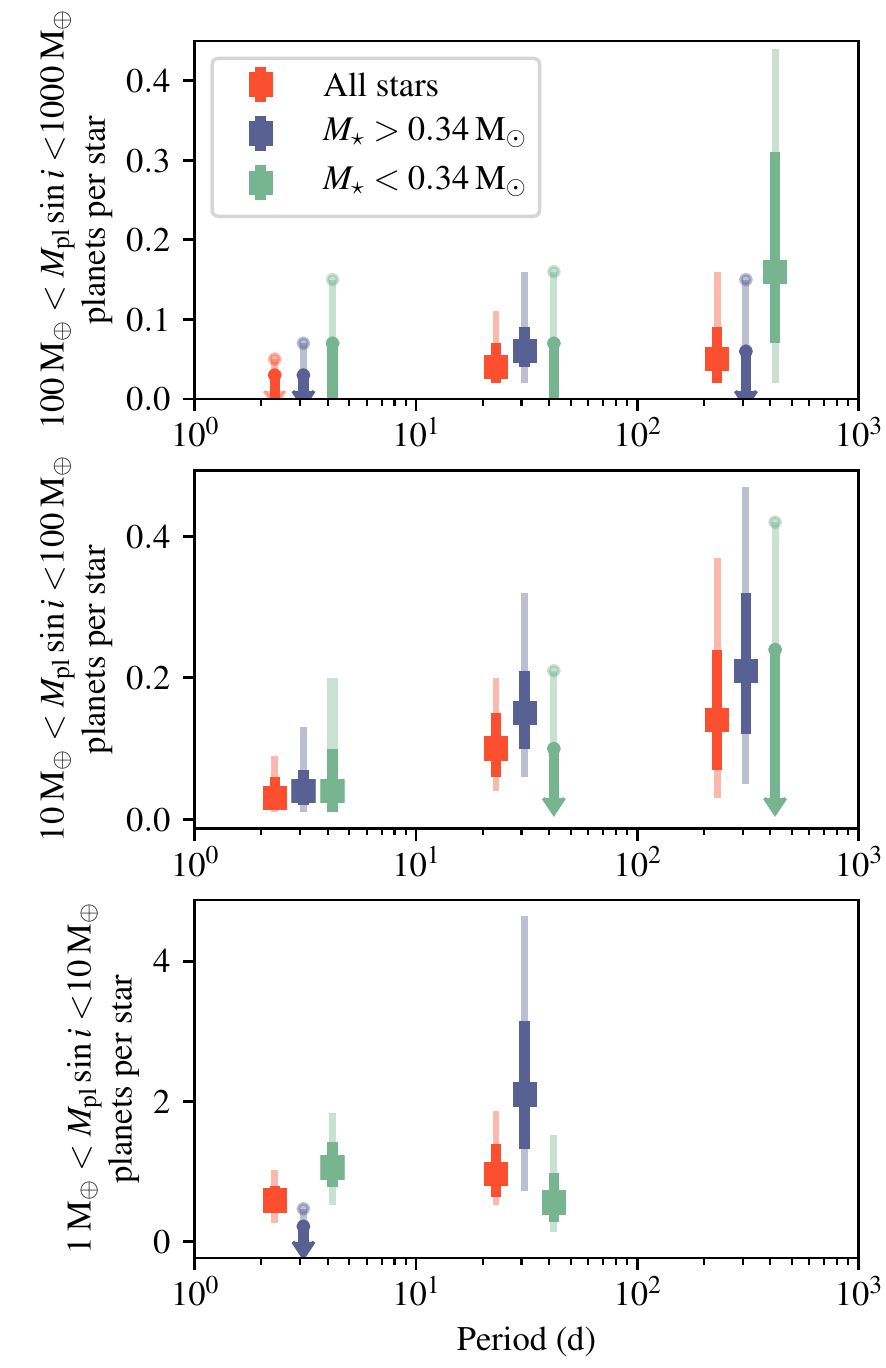}
 \caption{CARMENES M dwarf exoplanet occurrence rates as a function of orbital period for three different planetary mass intervals. Error bars are the 16\,\% and 84\,\% of the output distribution, and transparent error bars are the 2.5\,\% and 97.5\,\% of the output distributions. The occurrence rates of the upper panel might suffer from a selection bias (see Sect.~\ref{sec:discussion}).
 The colors have the same meaning as in Fig.~\ref{fig:sensitivity_mass}.}
 \label{fig:carmenes_occurrence}
\end{figure}

The minimum mass we gave every test planet in step 2 from above is drawn from this mass distribution. The resulting occurrence rates are shown in Fig.~\ref{fig:carmenes_occurrence}.

\subsection{Planets with $M_\text{pl} \sin i$ between \SI{100}{M_\oplus} and \SI{1000}{M_\oplus}}

\begin{table}
\caption{Planet occurrence rates (part 1). }
\label{table:split_mass_high_mass_planets}   
\centering 
\begin{tabular}{c c c}   
\hline\hline                
\noalign{\smallskip}
\multicolumn{3}{c}{\em (a) All M stars of the sample (71 stars)} \\
\multicolumn{3}{c}{$P$ (d)} \\
1--10 & 10--100 & 100--1000  \\  
\hline                                   
\noalign{\smallskip}
$N_\text{pl,det}$ =  0  & $N_\text{pl,det}$ =   2              & $N_\text{pl,det}$ =   1  \\  
 $\overline{n}_\text{pl} <  0.03 $ & $\overline{n}_\text{pl}$ = 0.04$^{+0.03}_{-0.02}$    & $\overline{n}_\text{pl}$ = 0.05$^{+0.04}_{-0.03}$  \\
\noalign{\smallskip}
\hline                                             \noalign{\smallskip}
\multicolumn{3}{c}{\em (b) M stars with $\si{M_\star} > \SI{0.34}{M_\odot}$ (48 stars)} \\
\multicolumn{3}{c}{$P$ (d)} \\
1--10 & 10--100 & 100--1000  \\   
\noalign{\smallskip}
\hline                               
\noalign{\smallskip}
$N_\text{pl,det}$ =  0     & $N_\text{pl,det}$ =   2             & $N_\text{pl,det}$ =   0     \\  
 $\overline{n}_\text{pl} <  0.04  $  & $\overline{n}_\text{pl}$ = 0.06$^{+0.04}_{-0.03}$   & $\overline{n}_\text{pl} < 0.07$         \\
\noalign{\smallskip}
\hline                                            
\noalign{\smallskip}
\multicolumn{3}{c}{\em (c) M stars with $\si{M_\star} < \SI{0.34}{M_\odot}$ (23 stars)} \\
\multicolumn{3}{c}{$P$ (d)} \\
 1--10 & 10--100 & 100--1000  \\  
\noalign{\smallskip}
\hline  
\noalign{\smallskip}
  $N_\text{pl,det}$ =  0              & $N_\text{pl,det}$ =   0           & $N_\text{pl,det}$ =   1 \\  
$\overline{n}_\text{pl} <  0.08$     & $\overline{n}_\text{pl} < 0.08$  & $\overline{n}_\text{pl}$ = 0.16$^{+0.15}_{-0.09}$       \\
\noalign{\smallskip}
\hline                                         
\end{tabular}
\tablefoot{
Planet occurrence rates of CARMENES planets in the minimum mass range of \SI{100}{M_\oplus} to \SI{1000}{M_\oplus}.\\
$\overline{n}_\text{pl}$: average number of planets per star;
$N_\text{pl,det}$: number of detected planets.
}
\end{table}

In Table~\ref{table:split_mass_high_mass_planets} and the upper panel of Fig.~\ref{fig:carmenes_occurrence}, we show the resulting occurrence rates for giant planets with $M_\text{pl} \sin i>\SI{100}{M_\oplus}$. As we do not detect any hot Jupiter ($P<10$\,d) in our sample, we place an upper bound on their occurrence rate at 0.03 planets per star. The most massive close-in planet that we find with CARMENES is GJ\,436\,b (originally discovered by \citealt{Butler2004} and reanalyzed by \citealt{Trifonov2018}), with a mass of \SI{21.4}{M_\oplus}. The paucity of hot Jupiters is also observed in all previous RV surveys of M dwarfs \citep{Endl2006, Zechmeister2009a, Bonfils2013}. The Exoplanet Encyplop{\ae}dia lists only four confirmed hot Jupiters around M dwarfs, all at very long heliocentric distances ($d> $ 140\,pc): \object{Kepler-45}\,b \citep{Johnson2012}, \object{HATS-6}\,b \citep{Hartman2015}, \object{NGTS-1}\,b \citep{Bayliss2018}, and \object{HATS-71}\,A\,b \citep{Bakos2020}. 

In comparison, the hot Jupiter occurrence rate from G-dwarf RV surveys is 1.1--1.6\,\% \citep[][]{Wright2012,Howard2010}. As our upper limit does not exclude this hot Jupiter frequency, a larger sample size is needed to confirm the lower occurrence rate of hot Jupiters around M dwarfs than around G dwarfs. At orbits from \SIrange{10}{1000}{d}, the giant planet fraction increases to 0.06$^{+0.04}_{-0.03}$ planets per star. The M dwarf giant planets reside in longer orbits and their overall fraction is comparable to that of gas giants around G dwarfs \citep[$5.2\,\%\pm0.6$\,\% in][]{Cumming2008}. However, this statement has to be taken with caution as the high frequency of gas giants in our sample is affected by a selection bias (see further details in Sect.~\ref{sec:discussion}). The frequency of M dwarf gas giants in the full sample with about five times as many targets could be up to a factor of five lower.

To explore the planet population dependence on host star mass, we split the sample into two groups of host star mass. We set the threshold between the groups at $M_\star = \SI{0.34}{M_\odot}$ (see Sect.~\ref{sec:completeness}). In the group of more massive, early M dwarfs there are 48 stars and in the group of less massive, late M dwarfs there are 23 stars. Within each subgroup, we repeated the occurrence rate calculations as described above. The resulting occurrence rates of both groups are also shown in Table~\ref{table:split_mass_high_mass_planets}. Due to the lower number of stars in each group, the uncertainties of those results are higher. We show detection maps of the stellar mass subsamples in Fig.~\ref{fig:carmenes_completeness_high_low_mass}. In the region of periods longer than 100\,d, our detection efficiency becomes low even for the high-mass planets. This is due to the strict period cutoff at half the time baseline of the observations. The median time baseline of the observations of our sample is 1124\,d. 

We detect only one giant planet around our less massive stars, which is the exceptional case of \object{GJ\,3512}\,b \citep{Morales2019}. This $\sim \SI{150}{M_\oplus}$ mass planet orbits a very low-mass star of only \SI{0.12}{M_\odot}. In addition, we confirm two giant planets around a host that is in our group of more massive stars: \object{GJ~876}\,b,c \citep{Trifonov2018}.

\subsection{Planets with $M_\text{pl} \sin i$ between \SI{10}{M_\oplus} and \SI{100}{M_\oplus}}

\begin{table}
\caption{Planet occurrence rates (part 2).}      \label{table:neptunes}     
\centering
\begin{tabular}{c c c}      
\hline\hline                     
\noalign{\smallskip}
\multicolumn{3}{c}{\em (a) All M stars of the sample (71 stars)} \\
 \multicolumn{3}{c}{$P$ (d)} \\
1--10 & 10--100 & 100--1000  \\   
\hline                                 
\noalign{\smallskip}
 $N_\text{pl,det}$ =  1                  & $N_\text{pl,det}$ =  5               & $N_\text{pl,det}$ =   2        \\ 
 $\overline{n}_\text{pl}$ = 0.03$^{+0.03}_{-0.01}$  & $\overline{n}_\text{pl}$ = 0.10$^{+0.05}_{-0.04}$    & $\overline{n}_\text{pl}$ = 0.14$^{+0.10}_{-0.07}$    \\
\noalign{\smallskip}
\hline                                             \noalign{\smallskip}
\multicolumn{3}{c}{\em (b) M stars with $\si{M_\star} > \SI{0.34}{M_\odot}$ (48 stars)} \\
 \multicolumn{3}{c}{$P$ (d)} \\
 1--10 & 10--100 & 100--1000  \\    
\noalign{\smallskip}
\hline                               
\noalign{\smallskip}
$N_\text{pl,det}$ =  1                  & $N_\text{pl,det}$ =   5             & $N_\text{pl,det}$ =   2             \\  
 $\overline{n}_\text{pl}$ = 0.04$^{+0.04}_{-0.02}$       & $\overline{n}_\text{pl}$ = 0.15$^{+0.07}_{-0.05}$   & $\overline{n}_\text{pl}$ = 0.21$^{+0.13}_{-0.10}$     \\
\noalign{\smallskip}
\hline                                            
\noalign{\smallskip}
\multicolumn{3}{c}{\em (c) M stars with $\si{M_\star} < \SI{0.34}{M_\odot}$ (23 stars)} \\
 \multicolumn{3}{c}{$P$ (d)} \\
 1--10 & 10--100 & 100--1000  \\  
\noalign{\smallskip}
\hline  
\noalign{\smallskip}
 $N_\text{pl,det}$ =  0              & $N_\text{pl,det}$ =   0           & $N_\text{pl,det}$ =   0     \\  
 $\overline{n}_\text{pl} < 0.10$    & $\overline{n}_\text{pl} < 0.10$                   & $\overline{n}_\text{pl} < 0.24$                  \\
\noalign{\smallskip}
\hline                                         
\end{tabular}
\tablefoot{
Planet occurrence rates of CARMENES planets in the minimum mass range of \SI{10}{M_\oplus} to \SI{100}{M_\oplus}. \\
$\overline{n}_\text{pl}$: average number of planets per star;
$N_\text{pl,det}$: number of detected planets.
}
\end{table}

In Table~\ref{table:neptunes} and the middle panel of Fig.~\ref{fig:carmenes_occurrence}, we show the resulting occurrence rates for planets with minimum masses from  \SIrange{10}{100}{M_\oplus}.
The occurrence rate of those intermediate-mass planets, including Saturns, Neptunes, and, perhaps, large super-Earths, increases from short-period orbits to long-period orbits. Again, we split the sample in the same stellar mass bins as described in the previous section. We find eight planets in this mass-period regime around the more massive stars, while we find no planets around our less massive stars. An analysis of the occurrence rates shows that from our upper limits we cannot tell if the population of intermediate-mass planets around our more massive stars is different from that of our less massive stars.

\subsection{Planets with $M_\text{pl} \sin i$ between \SI{1}{M_\oplus} and \SI{10}{M_\oplus}}\label{sec:low_mass_planets}

\begin{table}
\caption{Planet occurrence rates (part 3).}      \label{table:split_mass_low_mass_planets}   
\centering 
\begin{tabular}{c c c}         
\hline\hline                      
\noalign{\smallskip}
\multicolumn{3}{c}{\em (a) All M stars of the sample (71 stars)} \\
 \multicolumn{3}{c}{$P$ (d)} \\
 1--10 & 10--100 & 100--1000  \\   
\hline 
\noalign{\smallskip}
 $N_\text{pl,det}$ =  10     & $N_\text{pl,det}$ =   6      & $N_\text{pl,det}$ =   0       \\  
 $\overline{n}_\text{pl} = 0.59^{+0.20}_{-0.17}$        &  $\overline{n}_\text{pl} = 0.97^{+0.42}_{-0.33}$     &       ...        \\
\noalign{\smallskip}
\hline                                             \noalign{\smallskip}
\multicolumn{3}{c}{\em (b) M stars with $\si{M_\star} > \SI{0.34}{M_\odot}$ (48 stars)} \\
 \multicolumn{3}{c}{$P$ (d)} \\
 1--10 & 10--100 & 100--1000  \\ 
\noalign{\smallskip}
\hline                               
\noalign{\smallskip}
 $N_\text{pl,det}$ =  0 & $N_\text{pl,det}$ =   4            & $N_\text{pl,det}$ =   0 \\
$\overline{n}_\text{pl} < 0.22   $     & $\overline{n}_\text{pl}$ = 2.10$^{+1.13}_{-0.81}$  &  ...                \\
\noalign{\smallskip}
\hline                                            
\noalign{\smallskip}
\multicolumn{3}{c}{\em (c) M stars with $\si{M_\star} < \SI{0.34}{M_\odot}$ (23 stars)} \\
 \multicolumn{3}{c}{$P$ (d)} \\
 1--10 & 10--100 & 100--1000  \\   
\noalign{\smallskip}
\hline  
\noalign{\smallskip}
$N_\text{pl,det}$ =  10                  & $N_\text{pl,det}$ =   2                & $N_\text{pl,det}$ =   0  \\  
$\overline{n}_\text{pl}$ = 1.06$^{+0.35}_{-0.28}$       & $\overline{n}_\text{pl}$ = 0.55$^{+0.40}_{-0.26}$      &   ...                \\
\noalign{\smallskip}
\hline                                         
\end{tabular}
\tablefoot{
Planet occurrence rates of CARMENES planets in the minimum mass range of \SI{1}{M_\oplus} to \SI{10}{M_\oplus}. \\
$\overline{n}_\text{pl}$: average number of planets per star;
$N_\text{pl,det}$: number of detected planets.
}
\end{table}

The occurrence rates of our low-mass planets are shown in Table~\ref{table:split_mass_low_mass_planets} and the lower panel of Fig.~\ref{fig:carmenes_occurrence}. As expected, the low-mass planets ($M < \SI{10}{M_\oplus}$, i.e., Earths and super-Earths) are the most abundant type of planets in our sample. We detect ten planets close to their host star at periods less than 10\,d and six planets at intermediate periods of \SIrange{10}{100}{d}. The occurrence rates are also high with about 0.59 and 0.97 planets per star, respectively. There is a slight indication that the planet frequency increases toward long periods, but our results are also consistent with a flat occurrence rate distribution. From the results of transit surveys, we expect an overall higher number of low-mass planets around M dwarfs than around G dwarfs \citep[e.g.,][]{Howard2012, Mulders2015, Yang2020}. \cite{Mayor2011} derived a frequency of low-mass planets of $0.41\pm0.16$ planets per star with periods of up to 50\,d around G-type stars. In order to compare our results to this planet frequency, we ran our simulation again with the same period constraint.  The resulting low-mass planet frequency of our M dwarf sample is 1.18$^{+0.31}_{-0.27}$ planets per star. Our results, therefore, confirm a three times higher low-mass planet occurrence rate around M dwarfs.

We again computed the occurrence rates split in the two bins of stellar mass (see Table~\ref{table:split_mass_low_mass_planets}). All the low-mass planets of our sample with hosts with stellar masses of $M_\star>\SI{0.34}{M_\odot}$ reside in orbits longer than 10\,d. This means that in our earlier M dwarfs there is an increase in the low-mass planet occurrence rate toward long periods. In the lower mass stellar sample, on the other hand, there is a decreasing planet occurrence or a plateau toward long periods.

\section{Discussion}\label{sec:discussion}

\subsection{Assumptions and simplifications}\label{sec:makeitsimple}

To measure occurrence rates, we make several assumptions and simplifications, elaborated below.

\paragraph{No false positives.}
All planets included in this study are carefully analyzed. In the respective publications, the authors tested at least three activity indicators, orbit stability, and, in some cases, dynamical stability. All the planet signals have a very low FAP. Therefore, we assume that none of our planet candidates is a false positive. 

\paragraph{Unbiased sample.}
In any occurrence rate study, we need to be aware of the selection biases that could alter the statistics. 
When the original GTO sample was selected, the only selection criteria were spectral type, $J$-band magnitude, absence of known companions at less than 5\,arcsec, and visibility from the Calar Alto Observatory. After a few spectra were taken, new spectroscopic binaries were identified and excluded from the sample. This is usually done in RV surveys \cite[e.g., with HARPS and CORALIE by][]{Mayor2011} and is sometimes done in transit survey analysis as well \cite[e.g., in the \emph{Kepler} occurrence rate analysis of rocky habitable zone planets by][]{Bryson2021}. The results are most certainly different if close binary stars are kept in the sample \citep[e.g.,][]{Moe2021}. Therefore, this bias should be kept in mind if the results are compared to surveys that use other planet detection methods or artificial planet populations from planet formation theory. 
It should be noted that wide multiplicity does not affect our sample selection nor our RV survey \citep[e.g.,][]{Kaminski2018,Trifonov2018,GonzalezAlvarez2020}. 

In addition to that, our subsample of 71 stars out of the full GTO sample of 329 stars is not predefined. A high-mass planet can be identified with fewer observations than a low-mass planet. For this reason, planet detections of high-mass planets could be over-represented in the 71-star sample. If we do not find any additional giant planets in the rest of the 329 star sample, their occurrence rates will be up to a factor of five lower. An analysis of the full sample will give less biased giant planet occurrence rates.

\begin{figure}
    \centering
    \includegraphics[width=\linewidth]{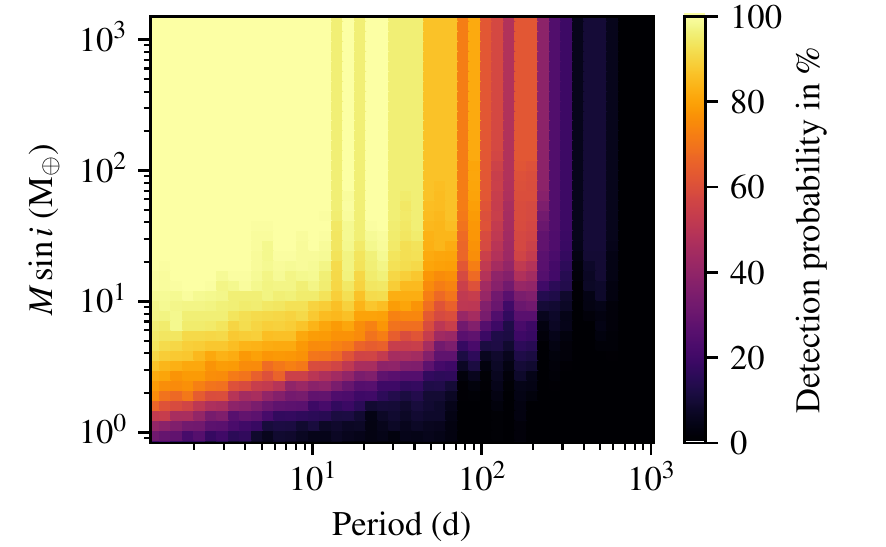}
    \caption{Same as Fig.~\ref{fig:carmenes_completeness} but for the subsample of 21 planet host stars.}
    \label{fig:carmenes_completeness_difference}
\end{figure}

Furthermore, many RV surveys observe targets with interesting signals more often than the rest of the sample. 
In this way, the detection limits of planet-hosting stars are at lower $M_\text{pl} \sin i$ than the detection limits of those without any planetary signal, as shown in Fig.~\ref{fig:carmenes_completeness_difference}. This could lead to an overestimation of the occurrence rates in the low-mass bins where the difference is the largest. We minimize this bias by already accepting test planets at a FAP as high as 1\,\% and averaging the occurrence rates over large bins of the $M_\text{pl} \sin i$-period plane.

\paragraph{Power-law distribution of injected $M_\text{pl} \sin i$.}
The power law that we determine as the underlying distribution is not very well constrained. 
Both power law fits are made to only three or four points. 
Still, this distribution is a much better approximation of the true underlying distribution than a log-uniform one. 
Choosing a realistic distribution from which to draw planets for the injection-recovery tests is especially important in the bins that contain large heterogeneities or gradients in detection sensitivity. 

As a comparison, in the lowest-mass bin with $\SI{1}{M_\oplus}< M_\text{pl} \sin i<\SI{10}{M_\oplus}$, we calculate planet occurrence rates of $0.35^{+0.12}_{-0.10}$ and $0.47^{+0.20}_{-0.15}$ planets per star for the 1--10\,d and the 10--100\,d bins, respectively, if we assume a log-uniform distribution of the injected $M_\text{pl} \sin i$. This is almost a factor of two lower than the rates that we obtain with a more realistic mass distribution.

\paragraph{Log-uniform distribution of period.}

Since the detection probability varies only weakly with period (e.g., 0--20\,\% in $P$ and 0--80\,\% in $M_\text{pl} \sin i$) within the chosen bins, the assumption of the distribution of $P$ is less critical compared to the one for $M_\text{pl} \sin i$.
A strong period dependence of the true occurrence rate, for which we see no evidence in our data, would thus not affect our analysis significantly.

\paragraph{Correct choice of method.}
We test other methods to determine detection limits and to retrieve planet occurrence rates. 
For the detection limits, we also use the method shown by  \cite{Howard2010}. 
They did not perform an injection-and-retrieval experiment but fit sinusoids to the data for a dense grid of orbital periods. 
The amplitude of the fit was considered as the detection limit. 
When we calculated the detection limits in this way, we retrieved detection limits that were up to a factor two lower than those calculated in Sect.~\ref{sec:completeness}. 
We consider the detection limits described in Sect.~\ref{sec:completeness} to be more realistic because the way with which we retrieve them is closer to the way with which we actually identify planet candidates. 
Nevertheless, the resulting occurrence rates are consistent in the high-mass bins. In the low-mass bins, the occurrences are lower (as expected with lower detection limits), but still consistent within the error bars. 

For retrieving planet occurrence rates, we also use the inverse detection efficiency method (IDEM), which is widely used in the literature \citep[e.g.,][]{Cumming2008, Wittenmyer2020}. 
The computed occurrence rates with this method are consistent in all period-mass bins with those described in Sect.~\ref{sec:occurrence}.
The main difference is that instead of an increasing occurrence rate with longer periods, we measure the same occurrence rate of $0.65\pm0.2$ planets per star in both low-mass bins with $M_\text{pl} \sin i < \SI{10}{M_\oplus}$. 
The reason for this is that this result is slightly dominated by four very low-mass planets around very low-mass stars (namely \object{Teegarden's Star b,c} and ; \citealt{Zechmeister2019}, \citealt{Stock2020}). 
Within IDEM, those planets get a much higher weight than the other planets. 
The results of this method, therefore, confirm the observational evidence of
the low-mass planets of low-mass stars residing in shorter-period orbits than those of stars with higher mass.

\paragraph{Circular orbits.} 
We fix the eccentricity of the Keplerian orbits to zero.
The consequence of this choice is explored in detail in Sect.~\ref{sec:eccen_multi}.

\paragraph{Separability.} 
We make use of the so-called approximation of separability \citep{Tremaine2012}: We treat multi-planet systems as several single-planet systems.
Again, this is explained in detail in Sect.~\ref{sec:eccen_multi}.

\subsection{Multi-planet systems and eccentric orbits}\label{sec:eccen_multi}

Our occurrence rate analysis is based on injection-and-retrieval experiments involving single-planets in circular orbits. This is of course an idealized setup, since many M dwarf planets reside in multi-planet systems \citep[e.g., TRAPPIST-1;][]{Gillon2017} and some are on eccentric orbits. Therefore, it is important to assess the impact of realistic planet multiplicity and orbital eccentricity distributions on our conclusions.

\begin{figure}
 \centering
\includegraphics[width=\linewidth]{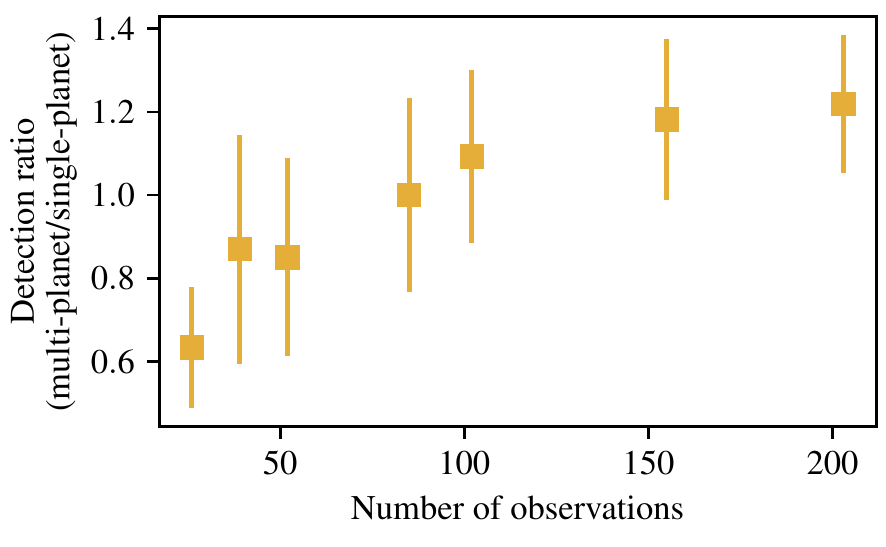}
\includegraphics[width=\linewidth]{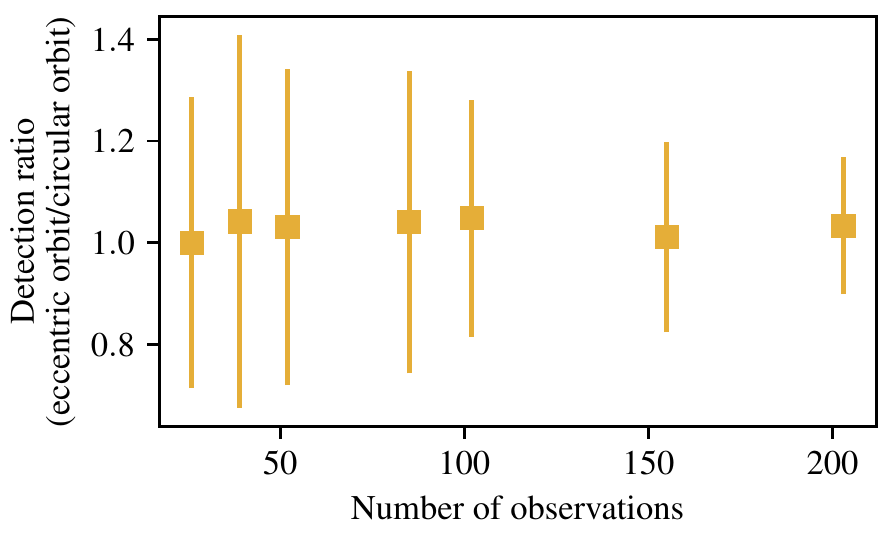}
\caption{Ratio of retrieved planets from injection-and-retrieval experiments with test planets in multi-planet test systems ({\em upper panel}) or on eccentric orbits ({\em lower panel}), compared to the same experiment with corresponding planets in single-planet systems or on circular orbits, respectively. Error bars are 16\,\% and 84\,\% of the simulation outcomes.}
 \label{fig:ecc_mult}
\end{figure}

The \enquote{approximation of separability} \citep{Tremaine2012} states that one can treat a multi-planet system like several single-planet systems with identical host stars. 
Therefore, we are interested in the ratio of detectable single-planet hosts in comparison to multi-planet hosts. If the approximation of separability is true, the $N_\text{hosts}$ that we can determine in a sample of multi-planet systems should be the same as that of a sample of identical single-planet systems that consist only of the planet of the first sample with the highest $K$ amplitude. We test this approximation with a set of artificial multi-planet systems. Our test systems are taken from a synthetic population created with the Generation 3 Bern Model \citep{Emsenhuber2020}, which was recently extended to M~dwarf hosts \citep{Burn2021}.
Almost all of the test systems are multi-planet systems with several planets. We ran an injection-and-retrieval experiment with those test planets. From the whole set of systems, we randomly drew 71 systems and calculated the corresponding RV curves. Our measurement errors and time stamps were taken from actual CARMENES observations of different stars with different numbers of RV values. The first simulation run was done on the whole test planet set. In a second run, we included only the planets with the highest RV semi-amplitude of each planetary system. The resulting ratio of retrieved planet hosts of multi-planet systems versus retrieved planet hosts of single-planet systems is plotted in the upper panel of Fig.~\ref{fig:ecc_mult}. In the case of only 26 RV values, we still retrieve about 60\,\% of the multi-planet hosts compared to the single-planet hosts. At the level of 50 observations, the ratio is 80\,\%. This number increases with the number of observations and, starting from about 80 observations, we retrieve the same number of multi-planet and single-planet hosts. 

The sensitivity of periodogram analysis is hardly affected by the orbital eccentricity as long as $e \leq 0.4$ \citep{Cumming2004}. \cite{Kipping2013} published an observed eccentricity distribution from known RV planets.
Of those planets, 80\,\% had an eccentricity of 0.4 or less. Nevertheless, we ran a simulation similar to that for the multi-planet systems. We took the single planets on eccentric orbits as the first set of test systems and the same planets on circular orbits as the second set of test systems. The result of the simulation is shown in the lower panel of Fig.~\ref{fig:ecc_mult}. The ratio is close to one even for a small number of observations.
For these reasons, the simplifications made in the injection-and-retrieval tests are not expected to bias our analysis statistically.

\subsection{Comparison to the HARPS M dwarf survey}\label{sec:harps}

\begin{figure}
 \centering
 \includegraphics[width=\linewidth]{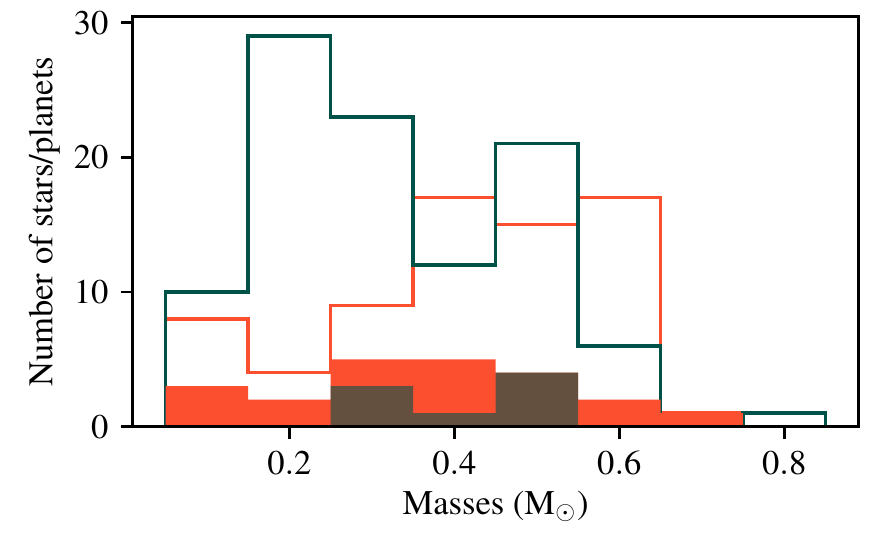}
  \includegraphics[width=\linewidth]{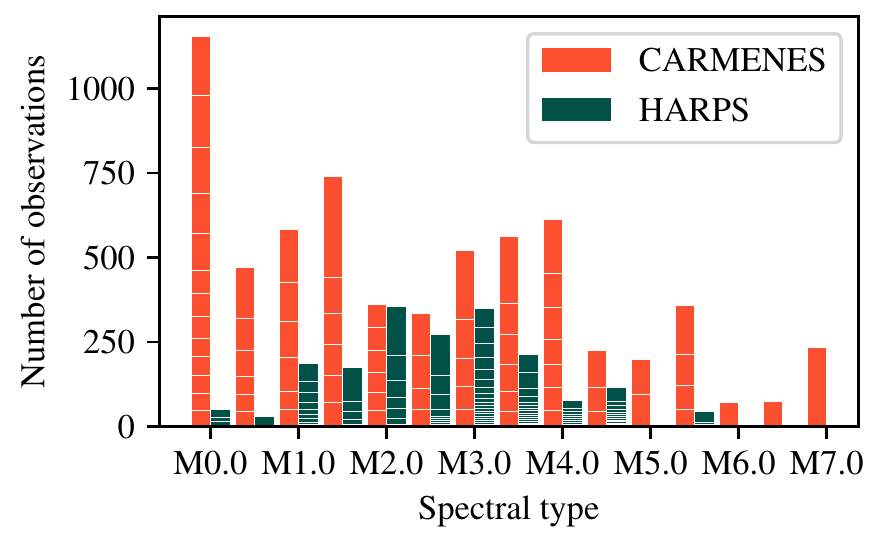}
 \caption{Comparison of the CARMENES and the HARPS M dwarf surveys. {\em Top panel}: Stellar mass distributions of the CARMENES (this work; red) and HARPS (\citealt{Bonfils2013}; dark green) surveys.
 Open histograms depict all target stars (CARMENES: 71; HARPS: 102).
 Filled histograms depict stars with planets (CARMENES: 22; HARPS:~8). 
 {\em Bottom panel}: Distribution of the number of spectra over spectral type in both surveys. HARPS spectral types are from \cite{Bonfils2013}, and the number of RV values until 1 April 2009
 is from \cite{Trifonov2020}.}
 \label{fig:harps_mass}
\end{figure}

\begin{figure}
 \centering
 \includegraphics[width=\linewidth]{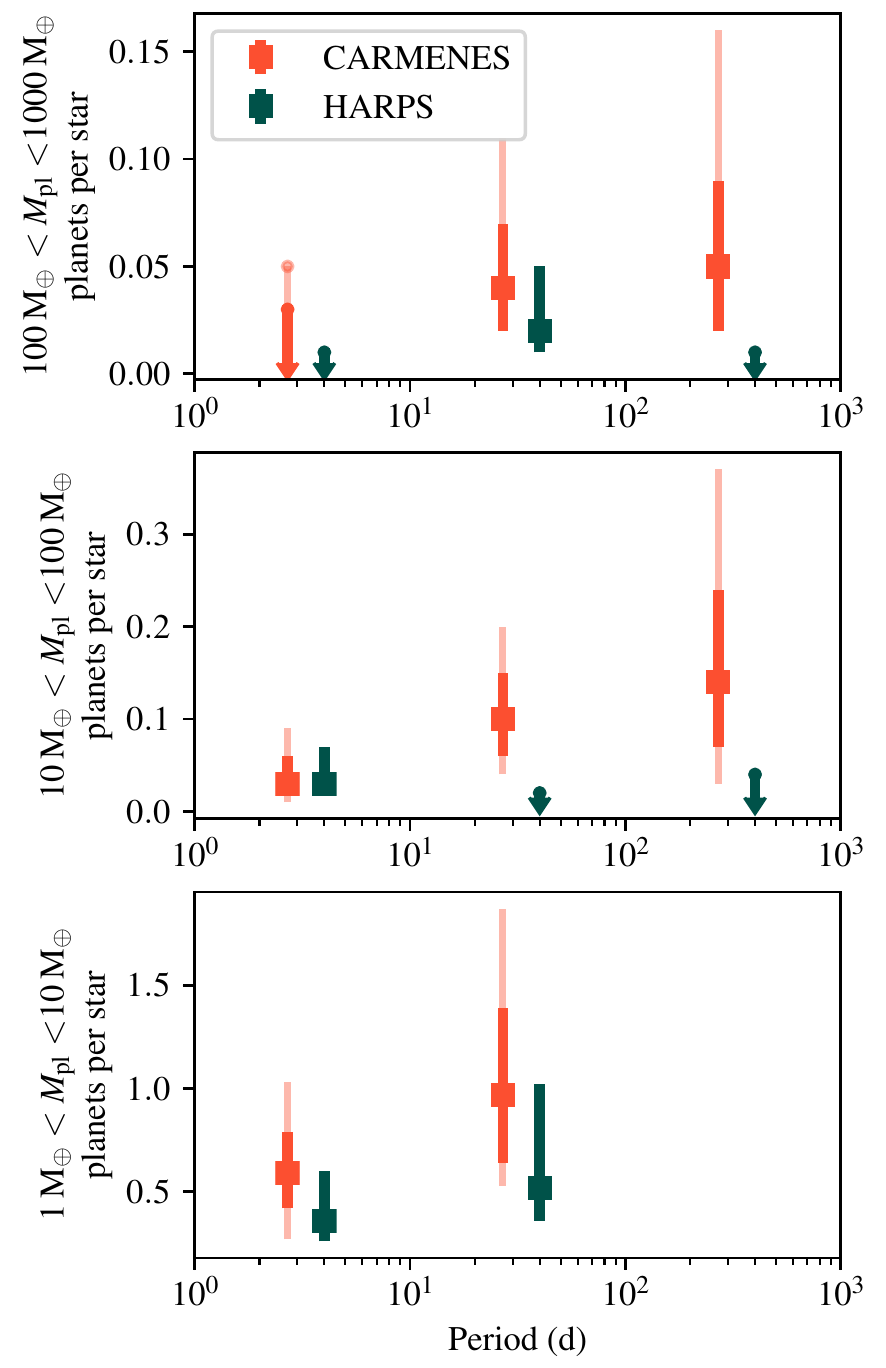}
 \caption{CARMENES (red) and HARPS (dark green) planet occurrence rates as a function of planet orbital period for three different planet mass intervals. The CARMENES occurrence rates of the upper panel might suffer from a selection bias (see Sect.~\ref{sec:discussion}).}
 \label{fig:harps_comparison}
\end{figure}

The largest previous RV study of M dwarfs is the HARPS M dwarf survey \citep{Bonfils2013}. 
It includes 102 stars and 14 reported planets\footnote{The HARPS planet sample includes the former planet candidate GJ\,581\,d, which is most probably an activity signal and, therefore, a false positive \citep{Robertson2014,Hatzes2016}.}. 
The extreme precision RV spectrograph HARPS covers the wavelength range \SIrange{378}{691}{nm} and, therefore, does not observe the range \SIrange{700}{900}{nm}, which is the sweet spot for M dwarf observations \citep{Reiners2018}. 
Nevertheless, the median stellar mass of the HARPS sample is \SI{0.29}{M_\odot} and, thus, lower than that of our CARMENES complete sample, which is \SI{0.43}{M_\odot} (see upper panel of Fig.~\ref{fig:harps_mass}). 
The median masses of the planet host stars in both surveys are very similar as well, with \SI{0.35}{M_\odot} and \SI{0.33}{M_\odot} for the CARMENES complete sample and the HARPS M dwarf sample, respectively. 
The main difference is that the CARMENES planet host stars have a wider range of masses from \SI{0.09}{M_\odot} to \SI{0.70}{M_\odot}, as compared to a range from \SI{0.31}{M_\odot} to \SI{0.49}{M_\odot} of the HARPS planet hosts (see also upper panel of Fig.~\ref{fig:harps_mass}). A comparison of the number of observations per spectral type of the two surveys (lower panel of Fig.~\ref{fig:harps_mass}) shows that the CARMENES observations are spread out more homogeneously over the spectral type range whereas the HARPS observations focus more on spectral types M1.0\,V-M4.0\,V.
A direct comparison of the \cite{Bonfils2013} occurrence rates with those of CARMENES is presented in Fig.~\ref{fig:harps_comparison}. The results are largely consistent. 
Only in the 10--100\,d and 10--100\,\si{M_\oplus} bin the CARMENES occurrence rate is higher. In this bin, CARMENES detected five planets, whereas \cite{Bonfils2013} reported zero detections. Two of the CARMENES planets are detected around M0.0\,V stars for which there are a lot more RV values from CARMENES than from HARPS. The upper limit of the HARPS data is lower than the 2.5\,\% output of our simulations (see Fig.~\ref{fig:harps_comparison}). It is possible that intermediate-mass planets are more frequent around earlier M dwarfs, as expected within the core accretion framework \cite[e.g.,][]{Burn2021}. In this case, the explanation for this is probably a combination of a better representation of early M dwarfs in the CARMENES measurements and a statistical effect. A statistical analysis with more data from the HARPS M dwarf sample or of the full CARMENES sample should show if this result remains significant.

\subsection{Comparison to small planets from \emph{Kepler}}\label{sec:Kepler}

\begin{figure}
 \centering
 \includegraphics[width=\linewidth]{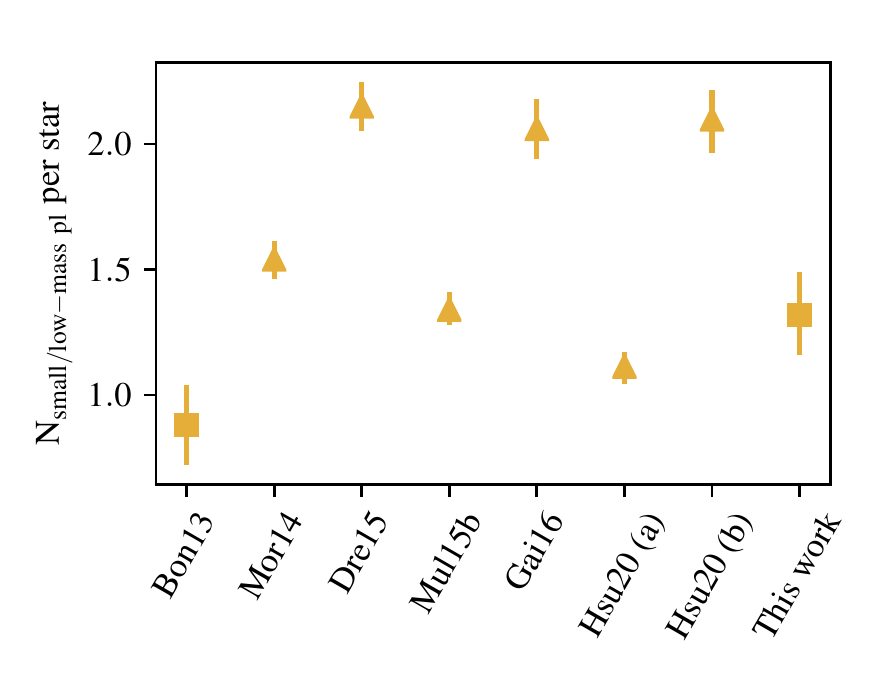}
 \caption{Small or low-mass planet occurrence rate in M dwarfs from different surveys in the ranges of 1--100\,d orbital period and \SIrange{1}{4}{R_\oplus} or \SIrange{1}{10}{M_\oplus}. Error bars are derived from the square root of the planets in the sample in all cases to make them comparable.
 Triangles show the results from the \emph{Kepler} transit surveys (\citealt{Morton2014}, \citealt{Dressing2015}, \citealt{Mulders2015a}, \citealt{Gaidos2016}, \citealt{Yang2020}, \citealt{Hsu2020}) and squares the results from the HARPS and CARMENES RV surveys \citep[][and this work]{Bonfils2013}.}
 \label{fig:comparison}
\end{figure}

Other widely used planet occurrence rates of M dwarfs are those derived from results of the {\em Kepler} space mission. The various studies that analyze the \emph{Kepler} M dwarf sample used subsamples of the one from \cite{Dressing2013}. In this sample of $\sim \num{4000}$ stars, about 20\,\% are in the mass range of our lower stellar mass sample (786 stars) but only 1.5\,\% are stars with less than \SI{0.15}{M_\odot} (58 stars). The median mass of this large sample is \SI{0.47}{M_\odot}, which is higher than the median mass of the sample, which is investigated in this paper (\SI{0.43}{M_\odot}). Therefore, we need to keep in mind that if there is a trend in occurrence rate with stellar mass, the occurrence rates are not exactly comparable.

The comparison with our occurrence rates is also not straightforward for a second reason: transit surveys probe the planet radius, $R_\text{pl}$, whereas RV surveys probe $M_\text{pl} \sin i$.
Mass and radius are related through bulk density, which can vary widely among small planets \citep[e.g.,][]{Hatzes2015, MartinezRodriguez2019}.
As an extreme example, two super-Earths orbiting the same star (K2--106, \object{TYC 608--458--1}) were found to have densities \SI{2.0}{g\,cm^{-3}} and \SI{13.1}{g\,cm^{-3}} \citep{Guenther2017}. 
Similar contrasting densities are also measured, with the help of RV measurements, in multi-planetary systems discovered with the {\em TESS}, such as \object{LTT 3780 b and c} \citep{Cloutier2020, Nowak2020}.
Therefore, the two parameters minimum mass and radius are not directly comparable. 
Nevertheless, we can compare the surveys on a statistical level with the mass-radius relations derived by \cite{Chen2017} for planets in general or by \cite{Kanodia2019} for M dwarf planets. 
According to these mass-radius relations, our mass bin of $M_\text{pl} \sin i$ and \SIrange{1}{10}{M_\oplus} roughly corresponds to the radius interval \SIrange{1}{4}{R_\oplus} in transit surveys. 
In Fig.~\ref{fig:comparison} we show a comparison of the low-mass or small planet occurrence rates in M dwarf samples as they are derived in various publications in the range of 1--100\,d. 
Our values for the low-mass planet frequency are consistent with what is found in some publications of \emph{Kepler} small planet occurrence rates, such as those by \cite{Mulders2015a} and \cite{Hsu2020}. Furthermore, the results by \cite{Morton2014}, \cite{Dressing2015}, \cite{Gaidos2016}, and \cite{Hsu2020} indicate a $\sim 1.5$ times higher small-planet occurrence rate than ours. \cite{Yang2020} report 2.1 planets per star around M dwarfs but their period and mass range is much broader than ours ($0.4-20$\,\si{R_\oplus} and $P_\text{pl} <400$\,d). In a smaller radius and period range, their result could be consistent with ours.  
In Fig.~\ref{fig:comparison}, we plot two occurrence rates from \cite{Hsu2020}, who work with two different distributions as priors for the planet size: (a) a Dirichlet prior over multiple radius bins per period range and (b) independent uniform priors for each bin.

\cite{HardegreeUllman2019} derive a small-planet occurrence rate as a function of spectral type (not shown in Fig.~\ref{fig:comparison}). 
They find that the small-planet frequency increases toward later spectral types. 
Their limits are radii from \SI{0.5}{R_\oplus} to \SI{2.5}{R_\oplus} at periods of less than 10\,d.
Therefore, their absolute numbers are not comparable to ours. Nevertheless, we would expect to see the very same trend in our data. 
The lower panel of Fig.~\ref{fig:carmenes_occurrence} shows exactly that behavior: in short orbits of up to 10\,d, the low-mass planet occurrence rate of our low-mass stars is significantly higher than that of our more massive stars.

\cite{Mulders2015} show that for various spectral types the small-planet frequency is lower at short orbital periods (in $\log P$ space). They obtain an occurrence rate of 27.9\,\% for orbital periods of 1--10\,d and 64.6\,\% for orbital periods of 10--100\,d. In a later publication based on more data from \emph{Kepler}, they revised those occurrence rates to 35.7\,\% and 98.3\,\%, respectively \citep{Mulders2015a}. Therefore, their short-period occurrence rates are 2.32 or 2.75 times lower than those in the longer-period orbits. We see the same trend in our data if we consider the whole stellar sample, although not as prominent. Our short-period occurrence rate is 1.64 times lower than that for longer-period orbits. The low-mass planet occurrence around low-mass stars, on the other hand, shows a reverse or flat trend in period. We want to know if our results could still be consistent with a 2.32 times lower occurrence rate in the short-period bin. We thus counted the cumulative probability of such a simulation outcome, which is 0.36\,\%. For this reason, we conclude that our results in the low stellar mass bin are not consistent with a drop in planet occurrence rate as high as 2.3 times. A reason for this could be the lower median stellar mass of our sample as compared to the sample in \cite{Mulders2015}. A more detailed study of planet occurrence rates for very low-mass stars will test if this result remains significant.

\section{Conclusions}\label{sec:conclusion}

We calculate preliminary planet occurrence rates from a first subsample of 71 CARMENES GTO M dwarfs. 
The giant planet ($\SI{100}{M_\oplus} < M_\text{pl} \sin i < \SI{1000}{M_\oplus}$) occurrence rate in our sample is 0.06$^{+0.04}_{-0.03}$\,\% for periods of up to 1000\,d. We set an upper limit of 0.03 planets per star for hot Jupiters ($P < 10$\,d). The most massive close-in planet that we detect has a mass of \SI{21.4}{M_\oplus}. 
Overall, the giant planet occurrence rate in our sample increases toward long periods, and it is lower than or equal to that around G-type stars. The analysis of the full 329 star sample of the CARMENES survey will show if there is a selection bias that could lead to an overestimation of the giant planet occurrence rate by up to a factor of five.

For intermediate-mass planets with minimum masses between \SI{10}{M_\oplus} and \SI{100}{M_\oplus}, we similarly find an increase in the planet frequency toward long periods, in contrast to the results presented by \cite{Bonfils2013}. The total occurrence rate is $0.18^{+0.07}_{-0.05}$ intermediate-mass planets per star.

Low-mass planets ($\SI{1}{M_\oplus} < M_\text{pl} \sin i < \SI{10}{M_\oplus}$) are very abundant in our sample: We measure an occurrence rate of $1.32^{+0.33}_{-0.31}$ low-mass planets per star for periods of up to 100\,d. This result is consistent with, or lower than, the results from the \emph{Kepler} survey (see Fig.~\ref{fig:comparison}). It confirms an at least twice higher abundance of low-mass planets around M dwarfs as compared to G dwarfs \citep[][ Sect.~\ref{sec:low_mass_planets}]{Mayor2011}. In a sample of late-type M dwarfs with stellar mass $M_\star <$ \SI{0.34}{M_\odot}, we find a very high low-mass planet occurrence rate in orbits shorter than 10\,d, which is in agreement with results from the \emph{Kepler} survey. Although from \emph{Kepler} results the planet occurrence rate is expected to be lower in short-period orbits, our results imply that the low-mass planet occurrence rate in longer-period orbits is the same as or lower than in shorter-period orbits around our sample of low-mass stars. 

We consolidate previous evidence for a high frequency of low-mass planets around the least massive stars, which poses constraints on the radial distribution and migration of planetary building blocks. 
Stellar mass-dependent planet formation models will have to explain the increased efficiency of turning these building blocks into planets in M dwarf systems.
To this end, an investigation of our findings with the core accretion model by~\citet{Burn2021} is already in progress (Schlecker et al., in prep.).
This and future comparisons between observed with theoretically predicted trends will help to shed light on different planet formation conditions as a function of stellar host mass.

\begin{acknowledgements}
We thank the anonymous referee for many useful comments and suggestions that helped improving our paper. We thank M.~Esposito for the explanation of his detection limit method and V.~Wolthoff for useful insights in occurrence rate retrieval methods.

This work was supported by the Th\"uringer Ministerium f\"ur Wirtschaft, Wissenschaft und Digitale Gesellschaft, and the Deutsche Forschungsgemeinschaft (DFG) under the DFG Research Unit FOR2544 \enquote{Blue Planets around Red Stars} (RE 2694/8-1) and the project GU~464/20-1.

CARMENES is an instrument at the Centro Astron\'omico Hispano-Alem\'an (CAHA) at Calar Alto (Almer\'{\i}a, Spain), operated jointly by the Junta de Andaluc\'ia and the Instituto de Astrof\'isica de Andaluc\'ia (CSIC).
  
  CARMENES was funded by the Max-Planck-Gesellschaft (MPG), 
  the Consejo Superior de Investigaciones Cient\'{\i}ficas (CSIC),
  the Ministerio de Econom\'ia y Competitividad (MINECO) and the European Regional Development Fund (ERDF) through projects FICTS-2011-02, ICTS-2017-07-CAHA-4, and CAHA16-CE-3978, 
  and the members of the CARMENES Consortium 
  (Max-Planck-Institut f\"ur Astronomie,
  Instituto de Astrof\'{\i}sica de Andaluc\'{\i}a,
  Landessternwarte K\"onigstuhl,
  Institut de Ci\`encies de l'Espai,
  Institut f\"ur Astrophysik G\"ottingen,
  Universidad Complutense de Madrid,
  Th\"uringer Landessternwarte Tautenburg,
  Instituto de Astrof\'{\i}sica de Canarias,
  Hamburger Sternwarte,
  Centro de Astrobiolog\'{\i}a and
  Centro Astron\'omico Hispano-Alem\'an), 
  with additional contributions by the MINECO, 
  the Deutsche Forschungsgemeinschaft through the Major Research Instrumentation Programme and Research Unit FOR2544 ``Blue Planets around Red Stars'', 
  the Klaus Tschira Stiftung, 
  the states of Baden-W\"urttemberg and Niedersachsen, 
  and by the Junta de Andaluc\'{\i}a.
  
  We acknowledge financial support from the Agencia Estatal de Investigaci\'on of the Ministerio de Ciencia, Innovaci\'on y Universidades and the ERDF through projects 
  PID2019-109522GB-C5[1:4]/AEI/10.13039/501100011033    
and the Centre of Excellence ``Severo Ochoa'' and ``Mar\'ia de Maeztu'' awards to the Instituto de Astrof\'isica de Canarias (SEV-2015-0548), Instituto de Astrof\'isica de Andaluc\'ia (SEV-2017-0709), and Centro de Astrobiolog\'ia (MDM-2017-0737), the Generalitat de Catalunya/CERCA programme,
the DFG program SPP 1992 \enquote{Exploring the Diversity of Extrasolar Planets} (JE 701/5-1), and
NASA (NNX17AG24G).

\end{acknowledgements}

\bibliographystyle{aa}
\bibliography{literature}

\begin{appendix}
\onecolumn
\section{Long table}
\tiny
\begin{longtable}{ll cc cc cc l}      
\caption{Output of the periodicity search program.}             
\label{table:planets}    \\
\hline\hline
Karmn & Name & $\alpha$ & $\delta$  & $M_\star$ & N$_\text{obs,VIS}$ & $P$  & FAP & Remark \\ 
~ & ~ & (J2016) & (J2016) & (\si{M_\odot}) & ~ & (d)  & (\%) & ~ \\ 
\hline
\endfirsthead\caption{continued.}\\
\hline\hline
Karmn & Name & $\alpha$ & $\delta$  & $M_\star$ & N$_\text{obs,VIS}$ & $P$  & FAP & Remark \\ 
~ & ~ & (J2016) & (J2016) & (\si{M_\odot}) & ~ & (d)  & (\%) & ~ \\ 
\hline
\endhead
J00051+457  &  GJ 2  &  1.3007478670  &  45.785915382  &  0.518  &  52  &  ...   &  ... &  No signal \\ 
J00067-075  &  GJ 1002  &  1.676464124  &  -7.546212317  &  0.116  &  89  &  21.17  &  0.521  &   Unsolved\\ 
J00183+440  &  GX And  &  4.612667737  &  44.024729578  &  0.391  &  114  &  40.65  &  0.585  &   Rotation\\ 
J01025+716  &  Ross 318  &  15.658240193  &  71.678173532  &  0.488  &  115  &  43.39  &  $< 10^{-4}$ &   CaIRT\\ 
J01026+623  &  BD+61 195  &  15.668732123  &  62.34543864  &  0.515  &  80  &  9.33  &  0.01  &   Rotation\\ 
J01026+623  &  BD+61 195  &  15.668732123  &  62.34543864  &  0.515  &  80  &  18.9  &  0.315  &   CaIRT, H$\alpha$\\ 
J01125-169  &  YZ Cet  &  18.133079241  &  -16.996243422  &  0.142  &  108  &  3.06  &  0.005  &   Planet\\ 
J01125-169  &  YZ Cet  &  18.133079241  &  -16.996243422  &  0.142  &  108  &  80.62  &  0.017  &   dLW\\ 
J01125-169  &  YZ Cet  &  18.133079241  &  -16.996243422  &  0.142  &  108  &  4.7  &  0.035  &   Planet\\ 
J02222+478  &  BD+47 612  &  35.562403121  &  47.88020753  &  0.551  &  47  &  28.23  &  0.007  &   CaIRT, dLW\\ 
J02362+068  &  BX Cet  &  39.071425223  &  6.877888387  &  0.262  &  50  &  ...   &  ... &  No signal \\ 
J02442+255  &  VX Ari  &  41.068748746  &  25.521801507  &  0.357  &  51  &  ...   &  ... &  No signal \\ 
J02530+168  &  Teegarden's Star  &  43.269144864  &  16.86490241  &  0.089  &  234  &  4.91  & $< 10^{-4}$ &   Planet\\ 
J02530+168  &  Teegarden's Star  &  43.269144864  &  16.86490241  &  0.089  &  234  &  11.41  & $< 10^{-4}$ &   Planet\\ 
J02530+168  &  Teegarden's Star  &  43.269144864  &  16.86490241  &  0.089  &  234  &  172.34  &  0.001  &   dLW\\ 
J03133+047  &  CD Cet  &  48.35301547  &  4.775188109  &  0.161  &  103  &  2.29  & $< 10^{-4}$ &   Planet\\ 
J03133+047  &  CD Cet  &  48.35301547  &  4.775188109  &  0.161  &  103  &  67.91  &  0.344  &   Rotation\\ 
J03463+262  &  HD 23453  &  56.585773281  &  26.214660371  &  0.562  &  50  &  ...   &  ... &  No signal \\ 
J04153-076  &  $o^{02}$ Eri C  &  63.829970524  &  -7.670429685  &  0.284  &  47  &  1.8  & $< 10^{-4}$ &   CRX\\ 
J04290+219  &  BD+21 652  &  67.250204922  &  21.923451047  &  0.650  &  150  &  12.53  &  0.005  &   Rotation\\ 
J04290+219  &  BD+21 652  &  67.250204922  &  21.923451047  &  0.650  &  150  &  25.07  &  0.004  &   CaIRT, dLW\\ 
J04290+219  &  BD+21 652  &  67.250204922  &  21.923451047  &  0.65  &  150  &  175.22  &  0.031  &   CRX\\ 
J04376+528  &  BD+52 857  &  69.422714035  &  52.89156892  &  0.578  &  119  &  16.3  &  0.193  &   CaIRT, dLW\\ 
J04376+528  &  BD+52 857  &  69.422714035  &  52.89156892  &  0.578  &  119  &  7.9  &  0.197  &   Unsolved\\ 
J04376+528  &  BD+52 857  &  69.422714035  &  52.89156892  &  0.578  &  119  &  422.79  &  0.219  &   Unsolved\\ 
J04588+498  &  BD+49 1280  &  74.711477815  &  49.848799932  &  0.589  &  55  &  8.97  &  0.008  &   Unsolved\\ 
J05314-036  &  HD 36395  &  82.867435093  &  -3.68623703  &  0.556  &  90  &  37.08  &  0.003  &   CaIRT, H$\alpha$, dLW\\ 
J05314-036  &  HD 36395  &  82.867435093  &  -3.68623703  &  0.556  &  90  &  10000.0  & $< 10^{-4}$ &   $P > $ time baseline/2\\ 
J06011+595  &  G 192-013  &  90.29510135819  &  59.593191709  &  0.257  &  79  &  83.39  &  0.07  &   dLW\\ 
J06011+595  &  G 192-013  &  90.29510135819  &  59.593191709  &  0.257  &  79  &  44.1  &  0.373  &   Unsolved\\ 
J06011+595  &  G 192-013  &  90.29510135819  &  59.593191709  &  0.257  &  79  &  21.52  &  0.663  &   Unsolved\\ 
J06103+821  &  GJ 226  &  92.584271112  &  82.101001876  &  0.415  &  57  &  10000.0  &  0.023  &   $P > $ time baseline/2\\ 
J06105-218  &  HD 42581 A  &  92.643599341  &  -21.867723301  &  0.528  &  51  &  2621.41  & $< 10^{-4}$ &   $P > $ time baseline/2\\ 
J06371+175  &  HD 260655  &  99.291542865  &  17.566269572  &  0.456  &  55  &  ...   &  ... &  No signal \\ 
J06548+332  &  Wolf 294  &  103.700249916  &  33.266463248  &  0.36  &  206  &  14.21  & $< 10^{-4}$ &   Planet\\ 
J06548+332  &  Wolf 294  &  103.700249916  &  33.266463248  &  0.36  &  206  &  67.59  & $< 10^{-4}$ &   Rotation\\ 
J06548+332  &  Wolf 294  &  103.700249916  &  33.266463248  &  0.36  &  206  &  119.48  & $< 10^{-4}$ &   Rotation\\ 
J08413+594  &  LP 090-018  &  130.331661582  &  59.491836367  &  0.123  &  146  &  206.39  & $< 10^{-4}$ &   Planet\\ 
J08413+594  &  LP 090-018  &  130.331661582  &  59.491836367  &  0.123  &  146  &  2236.05  & $< 10^{-4}$ &   $P > $ time baseline/2\\ 
J08413+594  &  LP 090-018  &  130.331661582  &  59.491836367  &  0.123  &  146  &  39.3  &  0.038  &   Unsolved\\ 
J09143+526  &  HD 79210  &  138.583915757  &  52.684159157  &  0.586  &  70  &  16.32  & $< 10^{-4}$ &   CaIRT, H$\alpha$, dLW\\ 
J09143+526  &  HD 79210  &  138.583915757  &  52.684159157  &  0.586  &  70  &  1468.72  &  0.02  &   $P > $ time baseline/2\\ 
J09144+526  &  HD 79211  &  138.591672199  &  52.683521065  &  0.592  &  153  &  1432.22  & $< 10^{-4}$ &   $P > $ time baseline/2\\ 
J09144+526  &  HD 79211  &  138.591672199  &  52.683521065  &  0.592  &  153  &  24.4  &  0.001  &   Planet\\ 
J09144+526  &  HD 79211  &  138.591672199  &  52.683521065  &  0.592  &  153  &  16.66  & $< 10^{-4}$ &   CaIRT, H$\alpha$, dLW\\ 
J09561+627  &  BD+63 869  &  149.033267235  &  62.785950488  &  0.574  &  67  &  18.66  & $< 10^{-4}$ &   CaIRT, dLW\\ 
J09561+627  &  BD+63 869  &  149.033267235  &  62.785950488  &  0.574  &  67  &  8.93  &  0.028  &   Unsolved\\ 
J10122-037  &  AN Sex  &  153.072958611  &  -3.746713797  &  0.526  &  73  &  10.65  & $< 10^{-4}$ &   Rotation\\ 
J10122-037  &  AN Sex  &  153.072958611  &  -3.746713797  &  0.526  &  73  &  21.4  &  0.006  &   CaIRT\\ 
J10289+008  &  BD+01 2447  &  157.228867166  &  0.837848462  &  0.426  &  67  &  305.89  &  0.017  &   Unsolved\\ 
J10482-113  &  LP 731-058  &  162.055102017  &  -11.342590886  &  0.117  &  75  &  1.52  &  0.07  &   Rotation\\ 
J10482-113  &  LP 731-058  &  162.055102017  &  -11.342590886  &  0.117  &  75  &  2.93  &  0.032  &   dLW\\ 
J10564+070  &  CN Leo  &  164.102166667  &  7.002194444  &  0.132  &  73  &  2.7  & $< 10^{-4}$ &   CRX, dLW\\ 
J10584-107  &  LP 731-076  &  164.615773701  &  -10.775499854  &  0.208  &  45  &  4.62  & $< 10^{-4}$ &   CRX\\ 
J11000+228  &  Ross 104  &  165.015742945  &  22.831741481  &  0.386  &  60  &  ...   &  ... &  No signal \\ 
J11033+359  &  Lalande 21185  &  165.830959676  &  35.948653033  &  0.354  &  297  &  12.94  & $< 10^{-4}$ &   Planet \\ 
J11033+359  &  Lalande 21185  &  165.830959676  &  35.948653033  &  0.354  &  297  &  1960.31  & $< 10^{-4}$ &   $P > $ time baseline/2\\ 
J11054+435  &  BD+44 2051A  &  166.342583333  &  43.530972222  &  0.372  &  108  &  1043.71  &  0.001  &   $P > $ time baseline/2\\ 
J11110+304W  &  HD 97101 B  &  167.763593814  &  30.443921508  &  0.538  &  48  &  ...   &  ... &  No signal \\ 
J11417+427  &  Ross 1003  &  175.432608072  &  42.751586224  &  0.354  &  76  &  41.28  & $< 10^{-4}$ &   Planet\\ 
J11417+427  &  Ross 1003  &  175.432608072  &  42.751586224  &  0.354  &  76  &  514.72  & $< 10^{-4}$ &   Planet\\ 
J11421+267  &  Ross 905  &  175.550536327  &  26.703066902  &  0.426  &  99  &  2.64  & $< 10^{-4}$ &   Planet\\ 
J11421+267  &  Ross 905  &  175.550536327  &  26.703066902  &  0.426  &  99  &  56.29  &  0.932  &   Unsolved\\ 
J11511+352  &  BD+36 2219  &  177.779128394  &  35.273104447  &  0.456  &  109  &  11.12  &  0.003  &   Rotation\\ 
J11511+352  &  BD+36 2219  &  177.779128394  &  35.273104447  &  0.456  &  109  &  25.5  &  0.52  &   Unsolved\\ 
J12123+544S  &  HD 238090  &  183.08863753  &  54.486153045  &  0.578  &  108  &  13.68  & $< 10^{-4}$ &   Planet\\ 
J12123+544S  &  HD 238090  &  183.08863753  &  54.486153045  &  0.578  &  108  &  107.28  &  0.455  &   Activity\\ 
J12312+086  &  BD+09 2636  &  187.813082661  &  8.808353196  &  0.55  &  50  &  ...   &  ... &  No signal \\ 
J12479+097  &  Wolf 437  &  191.98153096  &  9.749418089  &  0.306  &  47  &  1.47  &  0.001  &   Planet\\ 
J13229+244  &  Ross 1020  &  200.733535753  &  24.463942534  &  0.272  &  92  &  3.02  & $< 10^{-4}$ &   Planet\\ 
J13229+244  &  Ross 1020  &  200.733535753  &  24.463942534  &  0.272  &  92  &  87.38  &  0.165  &   CRX, dLW\\ 
J14257+236W  &  BD+24 2733A  &  216.434869646  &  23.61227393  &  0.602  &  64  &  ...   &  ... &  No signal \\ 
J14307-086  &  BD-07 3856  &  217.693284972  &  -8.647357533  &  0.63  &  94  &  249.07  &  0.353  &   Unsolved\\ 
J16167+672S  &  HD 147379  &  244.172568593  &  67.239204096  &  0.627  &  175  &  86.9  & $< 10^{-4}$ &   Planet\\ 
J16167+672S  &  HD 147379  &  244.172568593  &  67.239204096  &  0.627  &  175  &  361.2  & $< 10^{-4}$ &   CRX\\ 
J16167+672S  &  HD 147379  &  244.172568593  &  67.239204096  &  0.627  &  175  &  22.06  &  0.002  &   CaIRT, H$\alpha$, dLW\\ 
J16303-126  &  V2306 Oph  &  247.574827573  &  -12.667686638  &  0.294  &  93  &  1.26  &  0.453  &   Planet\\ 
J16303-126  &  V2306 Oph  &  247.574827573  &  -12.667686638  &  0.294  &  93  &  17.88  &  0.001  &   Planet\\ 
J17303+055  &  BD+05 3409  &  262.594829884  &  5.547444738  &  0.537  &  54  &  33.77  &  0.605  &   CaIRT, H$\alpha$, CRX, dLW\\ 
J17378+185  &  BD+18 3421  &  264.476487839  &  18.595950946  &  0.426  &  100  &  15.52  &  0.001  &   Planet\\ 
J17378+185  &  BD+18 3421  &  264.476487839  &  18.595950946  &  0.426  &  100  &  480.52  &  0.007  &   Activity\\ 
J17378+185  &  BD+18 3421  &  264.476487839  &  18.595950946  &  0.426  &  100  &  40.3  &  0.004  &   CaIRT, H$\alpha$\\ 
J17578+046  &  Barnard's Star  &  269.448614358  &  4.737980766  &  0.172  &  199  &  311.25  &  0.001  &   Rotation\\ 
J18051-030  &  HD 165222  &  271.284034579  &  -3.032751785  &  0.45  &  53  &  ...   &  ... &  No signal \\ 
J18174+483  &  TYC 3529-1437-1  &  274.354394178  &  48.367522828  &  0.587  &  69  &  16.04  &  0.324  &   Rotation\\ 
J18198-019  &  HD 168442  &  274.961836363  &  -1.93861271  &  0.593  &  136  &  ...   &  ... &  No signal \\ 
J19169+051N  &  V1428 Aql  &  289.227732239  &  5.163161488  &  0.484  &  123  &  104.24  & $< 10^{-4}$ &   Planet\\ 
J19169+051N  &  V1428 Aql  &  289.227732239  &  5.163161488  &  0.484  &  123  &  174.48  &  0.001  &   Activity\\ 
J19169+051N  &  V1428 Aql  &  289.227732239  &  5.163161488  &  0.484  &  123  &  23.67  &  0.498  &   CRX\\ 
J19346+045  &  BD+04 4157  &  293.668264631  &  4.583853199  &  0.564  &  49  &  2.52  &  0.643  &   Unsolved\\ 
J20305+654  &  GJ 793  &  307.63811581  &  65.450778811  &  0.385  &  53  &  ...   &  ... &  No signal \\ 
J20533+621  &  HD 199305  &  313.332456219  &  62.151065131  &  0.529  &  156  &  118.33  &  0.398  &   CRX\\ 
J20533+621  &  HD 199305  &  313.332456219  &  62.151065131  &  0.529  &  156  &  183.37  &  0.166  &   Unsolved\\ 
J21164+025  &  LSPM J2116+0234  &  319.114751335  &  2.580771066  &  0.43  &  81  &  14.45  & $< 10^{-4}$ &   Planet\\ 
J21164+025  &  LSPM J2116+0234  &  319.114751335  &  2.580771066  &  0.43  &  81  &  42.98  & $< 10^{-4}$ &   CaIRT\\ 
J21348+515  &  Wolf 926  &  323.712922362  &  51.53845905  &  0.446  &  70  &  26.34  &  0.332  &   Rotation\\ 
J21466+668  &  G 264-012  &  326.671959776  &  66.803848043  &  0.297  &  159  &  8.05  & $< 10^{-4}$ &   Planet\\ 
J21466+668  &  G 264-012  &  326.671959776  &  66.803848043  &  0.297  &  159  &  2.31  & $< 10^{-4}$ &   Planet\\ 
J21466+668  &  G 264-012  &  326.671959776  &  66.803848043  &  0.297  &  159  &  92.47  & $< 10^{-4}$ &   H$\alpha$\\ 
J22021+014  &  BD+00 4810  &  330.540864624  &  1.399031119  &  0.548  &  79  &  10.96  &  0.041  &   Unsolved\\ 
J22057+656  &  G 264-018 A  &  331.435726824  &  65.649665641  &  0.482  &  91  &  123.74  & $< 10^{-4}$ &   CRX\\ 
J22096-046  &  BD-05 5715  &  332.422993791  &  -4.640831742  &  0.468  &  59  &  2380.57  & $< 10^{-4}$ &   $P > $ time baseline/2\\ 
J22096-046  &  BD-05 5715  &  332.422993791  &  -4.640831742  &  0.468  &  59  &  10000.0  &  0.015  &   $P > $ time baseline/2\\ 
J22114+409  &  1RXS J221124.3+410000  &  332.850163961  &  40.999928455  &  0.16  &  53  &  ...   &  ... &  No signal \\ 
J22115+184  &  Ross 271  &  332.87688547  &  18.426973534  &  0.565  &  66  &  381.86  & $< 10^{-4}$ &   Unsolved\\ 
J22115+184  &  Ross 271  &  332.87688547  &  18.426973534  &  0.565  &  66  &  39.04  &  0.081  &   CaIRT, dLW\\ 
J22137-176  &  LP 819-052  &  333.432467485  &  -17.687062321  &  0.178  &  71  &  3.65  & $< 10^{-4}$ &   Planet\\ 
J22137-176  &  LP 819-052  &  333.432467485  &  -17.687062321  &  0.178  &  71  &  611.67  & $< 10^{-4}$ &   $P > $ time baseline/2\\ 
J22252+594  &  G 232-070  &  336.322145441  &  59.412502969  &  0.406  &  101  &  13.35  & $< 10^{-4}$ &   Planet\\
J22532-142  &  IL Aqr  &  343.323973712  &  -14.266595816  &  0.327  &  68  &  61.17  & $< 10^{-4}$ &   Planet\\ 
J22532-142  &  IL Aqr  &  343.323973712  &  -14.266595816  &  0.327  &  68  &  30.09  & $< 10^{-4}$ &   Planet\\ 
J23113+085  &  NLTT 56083  &  347.847583333  &  8.515583333  &  0.3  &  87  &  2225.31  & $< 10^{-4}$ &   $P > $ time baseline/2\\ 
J23113+085  &  NLTT 56083  &  347.847583333  &  8.515583333  &  0.3  &  87  &  141.09  & $< 10^{-4}$ &   Unsolved\\ 
J23216+172  &  LP 462-027  &  350.403614791  &  17.284434444  &  0.383  &  66  &  ...   &  ... &  No signal \\ 
J23351-023  &  GJ 1286  &  353.796960795  &  -2.392682457  &  0.118  &  71  &  ...   &  ... &  No signal \\ 
J23381-162  &  G 273-093  &  354.532727834  &  -16.236502771  &  0.374  &  55  &  ...   &  ... &  No signal \\ 
J23419+441  &  HH And  &  355.479993699  &  44.170596761  &  0.143  &  97  &  178.74  & $< 10^{-4}$ &   Unsolved\\ 
J23419+441  &  HH And  &  355.479993699  &  44.170596761  &  0.143  &  97  &  93.21  &  0.543  &   dLW\\ 
\hline          
\end{longtable}
\tablefoot{Output of the periodicity search program of the 71 stars of the complete CARMENES subsample. Up to three signals with a periodogram FAP $<$ 1\,\% are listed per star.}
\end{appendix}
\end{document}